\documentclass[12pt,a4paper]{article}
\usepackage{ytableau}
\usepackage{jheppub}
\usepackage{multirow}
\usepackage{chemarrow}
\usepackage{amsmath,amssymb,amsfonts,mathtools}
\usepackage{float}
\usepackage{subfig}
\usepackage{scalefnt}
\usepackage{wrapfig}
\usepackage{fancybox}
\usepackage{ifsym}
\usepackage{relsize}
\newcommand{\bea}{\begin{eqnarray}\displaystyle}
\newcommand{\eea}{\end{eqnarray}}
\newcommand{\nn}{\nonumber}
\newlength{\arrow}
{\setlength{\fboxsep}{15pt}
\setlength{\mylength}{\linewidth}%
\addtolength{\mylength}{-2\fboxsep}%
\addtolength{\mylength}{-2\fboxrule}%
\Sbox
\minipage{\mylength}%
\setlength{\abovedisplayskip}{0pt}%
\setlength{\belowdisplayskip}{0pt}%
\equation}%
{\endequation\endminipage\endSbox
\[\fbox{\TheSbox}\]}

\title{6d String Chains}
\author[1]{Abhijit Gadde,}
\author[2,3]{Babak Haghighat,}
\author[4]{Joonho Kim,}
\author[4]{Seok Kim,}
\author[2]{Guglielmo Lockhart,}
\author[2]{Cumrun Vafa}
\affiliation[1]{Institute for Advanced Study, Princeton, NJ 08540, USA}
\affiliation[2]{Jefferson Physical Laboratory, Harvard University, Cambridge, MA 02138, USA}
\affiliation[3]{Department of Mathematics, Harvard University, Cambridge, MA 02138, USA}
\affiliation[4]{Department of Physics and Astronomy and Center for Theoretical Physics, Seoul National University, Seoul 151-747, Korea.}

\abstract{We consider bound states of strings which arise in 6d (1,0) SCFTs that are realized in F-theory in terms of linear chains of spheres with negative self-intersections 1,2, and 4.  These include the strings associated
to $N$ small $E_8$ instantons, as well as the ones associated to M5 branes probing A and D type singularities in M-theory or D5 branes probing ADE singularities in Type IIB string theory.
We find that these bound states of strings admit (0,4) supersymmetric quiver descriptions and show how one can compute their elliptic genera.}

\begin{document}
\maketitle


\section{Introduction}

Recently considerable progress has been made in classifying and understanding basic properties of six-dimensional superconformal field theories \cite{Heckman:2013pva,Ohmori:2014pca,Ohmori:2014kda,
DelZotto:2014hpa,Heckman:2014qba,Intriligator:2014eaa,DelZotto:2014fia,
Heckman:2015bfa,Bhardwaj:2015xxa,DelZotto:2015isa,Miao:2015iba,Ohmori:2015pua, Haghighat:2013gba,Haghighat:2013tka,Haghighat:2014pva,Kim:2014dza,Haghighat:2014vxa}.  
Formulating these theories in a way which leads to a computational scheme of physical
amplitudes is still a major challenge.  One idea along these lines is to note that on their
tensor branch these theories have light strings as basic ingredients, which naturally suggests
that these strings should play a key role in any computational scheme.
For example, it is already known that their elliptic genus can be used to compute the superconformal index
of these 6d theories \cite{Lockhart:2012vp,Kim:2012qf,Haghighat:2013gba}. The main aim of this paper is to expand the known dictionary \cite{Haghighat:2013gba,Haghighat:2013tka,Kim:2014dza,Haghighat:2014vxa} for describing the degrees of freedom living on these strings.

The six-dimensional SCFTs that can be obtained via F-theory compactification were classified in \cite{Heckman:2013pva,DelZotto:2014hpa,Heckman:2015bfa} (see also \cite{Bhardwaj:2015xxa}). This includes all presently known 6d SCFTs, and it may well be the case that all consistent 6d SCFTs can be realized this way. In this context, six-dimensional SCFTs are realized by putting F-theory on elliptic Calabi-Yau threefold with a non-compact base $B$ with the property that all compact curves can be simultaneously contracted to zero size.
These curves are in fact  two-spheres that have negative self-intersection ranging
from 1 to 12 (excluding 9,10,11) and intersect one another at a point, and are arranged into trees according to specific rules. In particular, one can identify some basic building blocks (the curves with negative self-intersection $3,\dots, 8$ and 12, and a few combinations of two or three curves \cite{Morrison:2012np}) that can be glued together by joining them with $(-1)$-curves. Each curve gives rise to a tensor multiplet in the 6d theory. D3 branes wrapping a curve correspond to strings coupled to the corresponding tensor multiplet. Furthermore, gauge multiplets arise whenever the elliptic fiber degenerates over a curve, the nature of the resulting gauge group being determined by the type of degeneration. In particular, non-Higgsable clusters necessarily support a nontrivial gauge group, whose rank can be made larger by increasing the degree of the singularity.

Under favorable circumstances, bound states of strings admit a description in terms of a two-dimensional (0,4) quiver gauge theory. In this paper we extend the list of theories for which such a description is available to three additional classes of 6d SCFTs:
\begin{itemize}
\item The theory of M5 branes probing a singularity of A or D type. The first case was studied in \cite{Haghighat:2013tka} and corresponds to a linear chain of $(-2)$ curves and leads to a 6d theory with SU-type gauge group; the second case is novel and corresponds to a linear chain of alternating $(-1)$ and $(-4)$ curves which support respectively gauge group $SO(8+2p)$ and $Sp(p)$.
\item The theory of $ N $ small $ E_8 $ instantons, or equivalently $ N $ M5 branes probing the M9 plane of Ho\v rava-Witten theory \cite{  Seiberg:1996vs,Ganor:1996mu}. This corresponds to a single $(-1)$ curve linked to a chain of $ N - 1 $ curves of self-intersection $ -2 $. Upon circle compactification, this theory admits deformation by a parameter corresponding to the mass of a 5d anti-symmetric hypermultiplet. We focus on the case where this parameter is turned off.
\item The theory of $ N  $ D5 branes probing an ADE singularity. This corresponds in F-theory to an ADE configuration of $ -2 $ curves supporting gauge groups of $SU$ type.
\end{itemize}

Once the quiver gauge theory corresponding to a given configuration of strings is specified, the elliptic genus can be computed by means of localization \cite{Gadde:2013dda, Benini:2013xpa}. We do this for certain specific bound states of strings for the first two classes of theories, and for arbitrary bound states of strings for the third class.

The organization of this paper is as follows: In Section 2 we review some of the building blocks for the (0,4) supersymmetric 2d quiver gauge theories which will be needed for the description of the worldsheet degrees of freedom of the tensionless strings. In Section 3 we discuss the 2d quiver for the strings of the theory of M5 branes probing A- or D- type singularities.  In Section 4 we study the quiver for the strings of the theory of $N$ small $ E_8 $ instantons. In Section 5 we discuss the quiver for strings of the theory of D5 branes probing an ADE singularity.  In Appendix A we discuss a candidate 2d quiver for the strings of the theory of  $N$ small $E_8$ instantons  with mass deformation.

\section{Chains of Strings}
\label{sec:3}

We are interested in computing the elliptic genera of the strings that arise on tensor branches of 6d $(1,0)$ SCFTs with several tensor multiplets, along the lines of \cite{Haghighat:2014vxa}, and are wrapped around a torus of complex modulus $ \tau $. In this paper we aim to obtain 2d quiver gauge theories for a variety of  6d SCFTs that arise within M- and F-theory. These will generally consist of $(0,4)$ quiver gauge theories with gauge group $\prod_{i=1}^N G_i(k_i)$, where $G_i$ is the gauge group associated to $ k_i $ strings of the $i$th type, and will capture the dynamics of a bound state of such a collection of strings. The gauge groups arising in the theories discussed in this paper are either unitary, symplectic, or orthogonal. \\

Let us now discuss global symmetries of the 2d $(0,4)$ theory. A number of the global symmetries have a geometric origin, since they arise from rotations of an $\mathbb{R}^4_{||}$ along the worldvolume of the six-dimensional theory but transverse to the string's worldsheet, or from rotations of an $\mathbb{R}^4_{\perp}$ perpendicular to the six-dimensional worldvolume. Overall, an $SU(2)^4$ group acting as rotations of $ \mathbb{R}^4_{||}\times \mathbb{R}^4_\perp $:
\begin{eqnarray}
	\mathbb{R}^4_{||} & : & SU(2)_{A'} \times SU(2)_{\tilde A'}~, \nonumber \\
	\mathbb{R}^4_{\perp} & : & SU(2)_A \times SU(2)_Y.
\end{eqnarray}
The right-moving supercharges $Q_+^{AA'}$ transform under the $SO(4)$ R-symmetry given by
\begin{equation}
	SO(4)_R = SU(2)_A \times SU(2)_{A'}.
\end{equation} 

Let us identify the Cartan of $ SU(2)^4 $ as follows:
\[ U(1)_{\epsilon_+}\times U(1)_{\epsilon_-}\times U(1)_m\times U(1)_R \subset SU(2)_{A'}\times SU(2)_{\widetilde A'}\times SU(2)_Y\times SU(2)_A,  \]
where we identify $U(1)_R$ with the R-symmetry group of the the $(0,4)$ theory when viewed as a $(0,2)$ theory. When computing elliptic genera, we will turn on fugacities $ \epsilon_+ = \frac{1}{2}(\epsilon_1+\epsilon_2), \epsilon_- = \frac{1}{2}(\epsilon_1-\epsilon_2),$ and $ m $ for the remaining $ U(1) $ factors in the Cartan.

Below we summarize our notation for the following sections. In drawing quiver diagrams, we will denote $(0,2)$ Fermi multiplets by dashed lines and $(0,4)$ hypermultiplets by a solid line. The various fields of the quiver gauge theory can be organized in terms of four combinations of (0,2) multiplets:
\begin{enumerate}
\item To each gauge node $ i $ corresponds the following field content valued in representations of $ G_i$ (corresponding to $ n_i $ strings of the $ i $th kind): a vector multiplet $ \Upsilon_i $; a Fermi multiplet $ \Lambda_i^\Phi $; and two chiral multiplets $ B_i,\widetilde B_i $.
\begin{center}
\begin{tabular}{|c|c|c|c|c|c|c|c|}\hline
Symbol & (0,2) field content & $U(1)_{\epsilon_1}$ & $U(1)_{\epsilon_2}$ & $U(1)_m$ & $U(1)_R$ & $G_i$\\
 \hline
 \multirow{4}{*} {\vspace{-.1in}\includegraphics[width=0.9in]{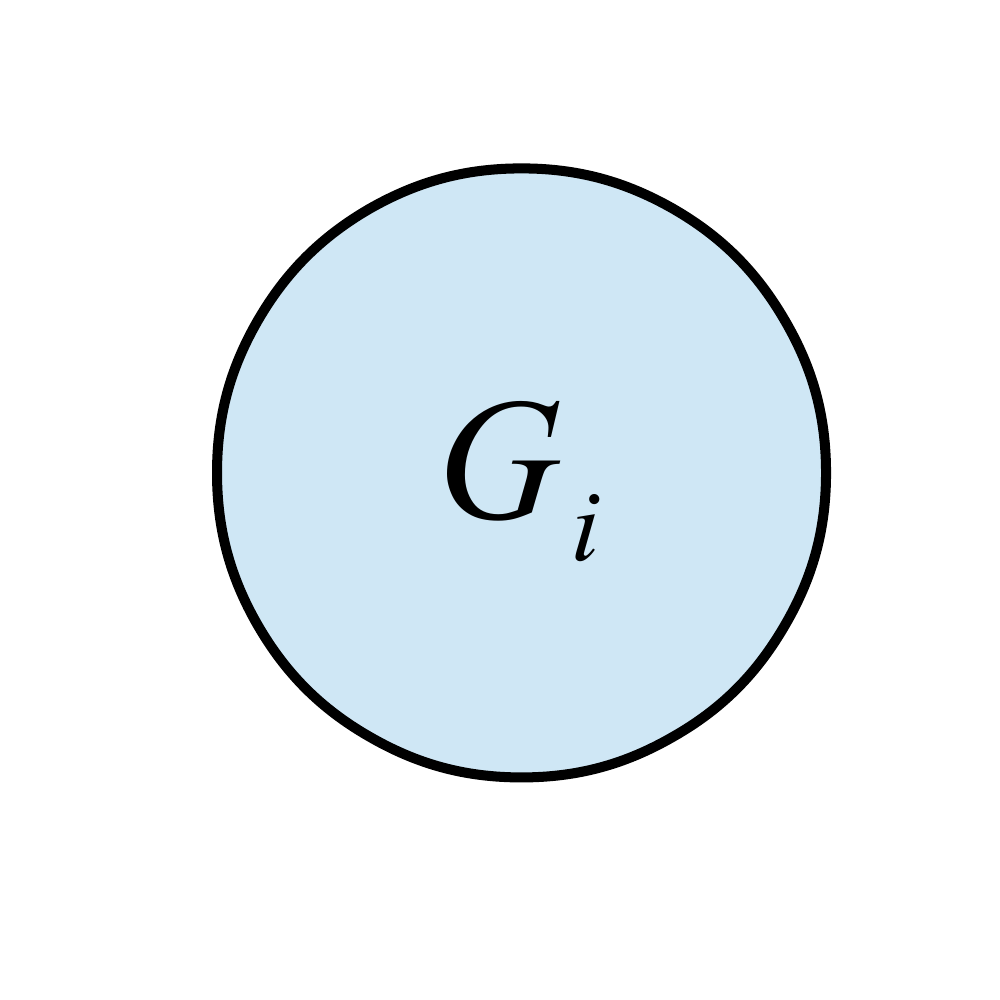}} & $ \Upsilon_i $ (vector) & 0 & 0 & 0 & 0 & \textbf{adj.}\\
 &$\Lambda_i^\Phi$ (Fermi) & $-1$ & $-1$ & 0 & 1 & \textbf{adj.}\\
 & $B$ (chiral) & 1 & 0 & 0 & 0 & \textbf{R}\\
 & $\widetilde B$ (chiral) & 0 & 1 & 0 & 0 & \textbf{R}\\
 \hline
\end{tabular}
\end{center}
The representation \textbf{R} is the adjoint representation whenever the gauge group is unitary, symmetric whenever the gauge node is orthogonal, and anti-symmetric if the gauge group is symplectic.
\item Between each pair of nodes $ i,j $ such that the corresponding element of the adjacency matrix of the underlying quiver $ M_{ij} $ is non-zero one has the following bifundamental fields of $ G_i\times G_j $: two Fermi multiplets $ \Lambda^B_{ij}, \Lambda^{\widetilde B}_{ij} $,  and chiral multiplets $\Phi_{ij} , \Sigma_{ij}$ forming a twisted $(0,4)$ hypermultiplet.  
\begin{center}
\begin{tabular}{|c|c|c|c|c|c|c|c|}\hline
Symbol & (0,2) field content & $U(1)_{\epsilon_1}$\!\! & \!$U(1)_{\epsilon_2}$\!\! & $U(1)_m$\!\! & \!$U(1)_R$\!\! & $G_i\!\!\times\!\!G_j$\\
 \hline
 \multirow{4}{*} {\vspace{-.1in}\includegraphics[width=.6in]{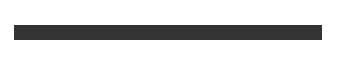}} 
 & $ \Lambda_{ij}^B $ (Fermi) & 1/2 & $-1/2$ & 1 & $0$ & $\square\otimes\overline\square$\\
 &$\Lambda_{ij}^{\widetilde B}$ (Fermi) & $-1/2$ & 1/2 & 1 & $0$ & $\square\otimes\overline\square$\\
 & $\Sigma_{ij}$ (chiral) & $-1/2$ & $-1/2$ & $1$ &1 & $\square\otimes\overline\square$\\
 &$\Phi_{ij} $  (chiral) & $-1/2$ & $-1/2$ & -1 & 1 & $\overline\square\otimes\square$\\
 \hline
\end{tabular}
\end{center}

\item Between each gauge node $ i $ and the corresponding global symmetry node one has a link corresponding to two chiral multiplets, $ Q,\widetilde Q $, charged under $ G_i \times F_i  $, where $ F_i $ is the global symmetry group at that node, which we depict by a square in the quiver.

\begin{center}
\begin{tabular}{|c|c|c|c|c|c|c|c|c|}\hline
Symbol & (0,2) field content & $U(1)_{\epsilon_1}$\!\! & \!$U(1)_{\epsilon_2}$\!\! & $U(1)_m$\!\! & \!$U(1)_R$\!\! & \!$G_i\!\!\times\!\! F_i$\\
 \hline
 \multirow{4}{*} {\vspace{-.1in}\includegraphics[width=.15in]{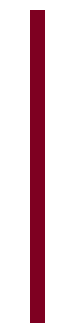}} &&&&&&\\
 & $Q_i$ (chiral) & 1/2 & 1/2 & 0 & 0 & $\square\otimes\overline\square$\\
 & $\widetilde Q_i $ (chiral) & 1/2 & 1/2 & 0 & 0 & $\overline\square\otimes\square$\\
&&&&&&\\
 \hline
\end{tabular}
\end{center}

\item Between each gauge node $ i $ and any successive node $ j $  one has a Fermi multiplet $ \Lambda_{ij}^Q $; between the same gauge node $ i $ and any preceding node $ j $  one has a Fermi multiplet $ \Lambda_{ji}^{\widetilde Q} $.

\begin{center}
\begin{tabular}{|c|c|c|c|c|c|c|c|c|}\hline
Symbol &\!\! (0,2) field content \!\!\!&\!\!\! $U(1)_{\epsilon_1}$\!\! &\!\!\! $U(1)_{\epsilon_2}$\!\! &\!\!\! $U(1)_m$\!\! &\!\!\! $U(1)_R$ \!\!\!&\!\!\! $G_i\!\! \times \!\!F_j$\!\\
 \hline
 \multirow{4}{*} {\vspace{-.1in}\includegraphics[width=.6in]{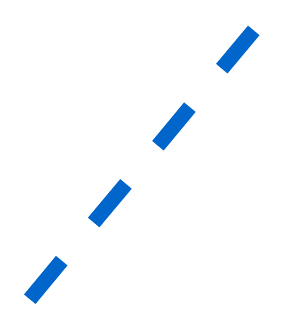}} &&&&&&\\
 & $\Lambda^Q_{ij}$ (chiral) & 0 & 0 & 1 & $0$ & $\square\otimes\overline\square$\\
 & $\Lambda^{\widetilde Q}_{ji} $ (chiral) & 0 & 0 & 1 & $0$ & $\overline\square\otimes\square$\\
&&&&&&\\
 \hline
\end{tabular}
\end{center}

\end{enumerate}

\section{Partition functions of M5 branes probing ADE Singularities}
\label{sec:1}

In this section we consider 6d $(1,0)$ SCFTs which arise from M5 branes probing singularities of type A and D, and obtain the 2d quiver gauge theory describing the self-dual strings that arise on the tensor branch of the corresponding 6d theory. 

\subsection{M5 branes probing an $A_{N-1}$ singularity}

Consider a setup where $M$ parallel M5 branes span directions $X^0,\dots, X^5$ and are separated along the $X^6$ direction in 11d spacetime. Taking the transverse space of the M5 branes to be $\mathbb{R}\times \mathbb{C}^2/\mathbb{Z}_N$ and blowing up the singular locus gives rise to a $(1,0)$ 6d SCFT on the tensor branch which enjoys a $SU(N)\times SU(N)$ flavor symmetry. The $A_{N-1}$ singularity can be thought of as a limit of Taub-NUT space with charge $N$; this space has a canonical circle fibration over $\mathbb{R}^3$, and compactifying M-theory along this circle one arrives at a system of $N$ parallel D6-branes stretched between NS5 branes. The dynamics of the strings that arise in this system are captured by the two-dimensional quiver theory of Figure \ref{fig:SuChain}:

\begin{figure}[h!]
  \centering
	\includegraphics[width=\textwidth]{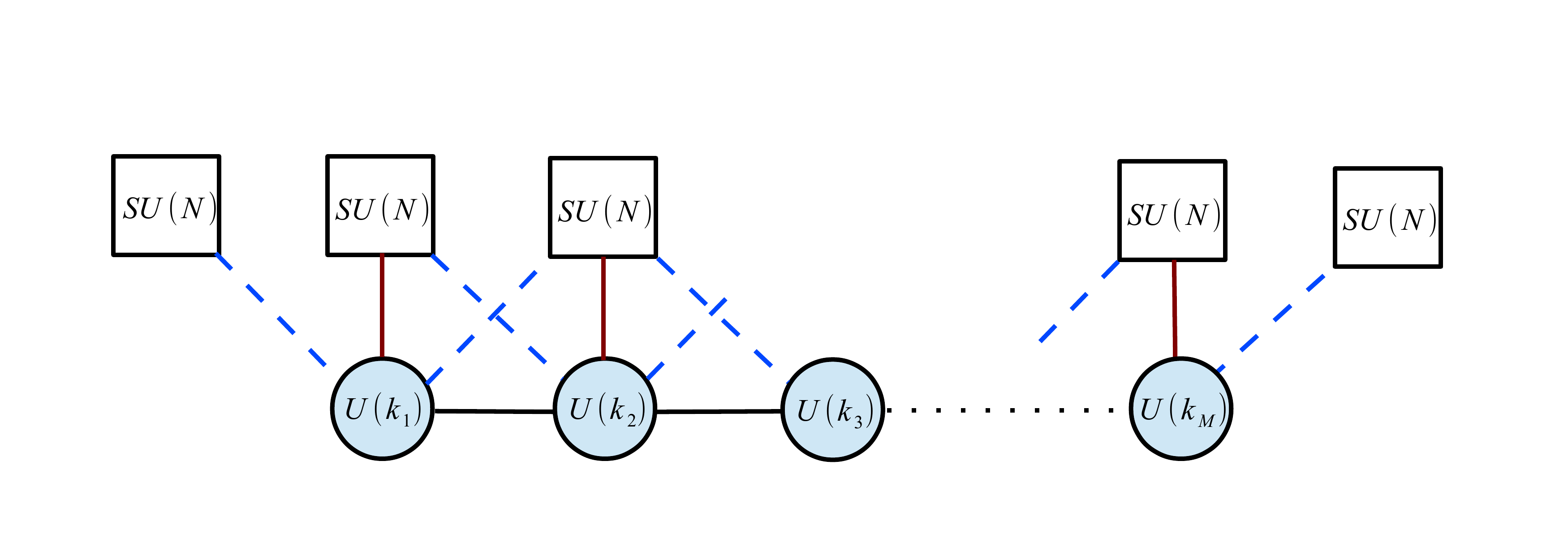}
  \caption{Non-critical strings in M5 branes probing $A_{N-1}$ singularities.}
  \label{fig:SuChain}
\end{figure}

The quiver corresponds to a 2d $\mathcal{N}=(0,4)$ theory obtained from a $\mathbb{Z}_M$ orbifold of a $\mathcal{N}=(4,4)$ supersymmetric Yang-Mills theory. As such, each gauge node contains a $(0,4)$ vector multiplet together with an adjoint $(0,4)$  hypermultiplet and the bifundamental fields between the gauge nodes consist of $(0,4)$ Fermi and twisted hypermultiplets. Furthermore, between each gauge node and the adjacent flavor nodes one has $(0,2)$ Fermi multiplets in the fundamental representation of the gauge group and between each gauge node and the corresponding flavor node a fundamental $(0,4)$ hypermultiplet. The exact field content is described in \cite{Haghighat:2013tka}. Following the rules of \cite{Benini:2013xpa} and the charge tables of Section \ref{sec:3} one can straightforwardly write down an expression for the elliptic genus for any configuration of strings, corresponding to different choices of the ranks of the gauge groups in the 2d quiver. We will perform the computation in Section \ref{sec:D5}.

\begin{figure}[h!]
  \centering
	\includegraphics[width=\textwidth]{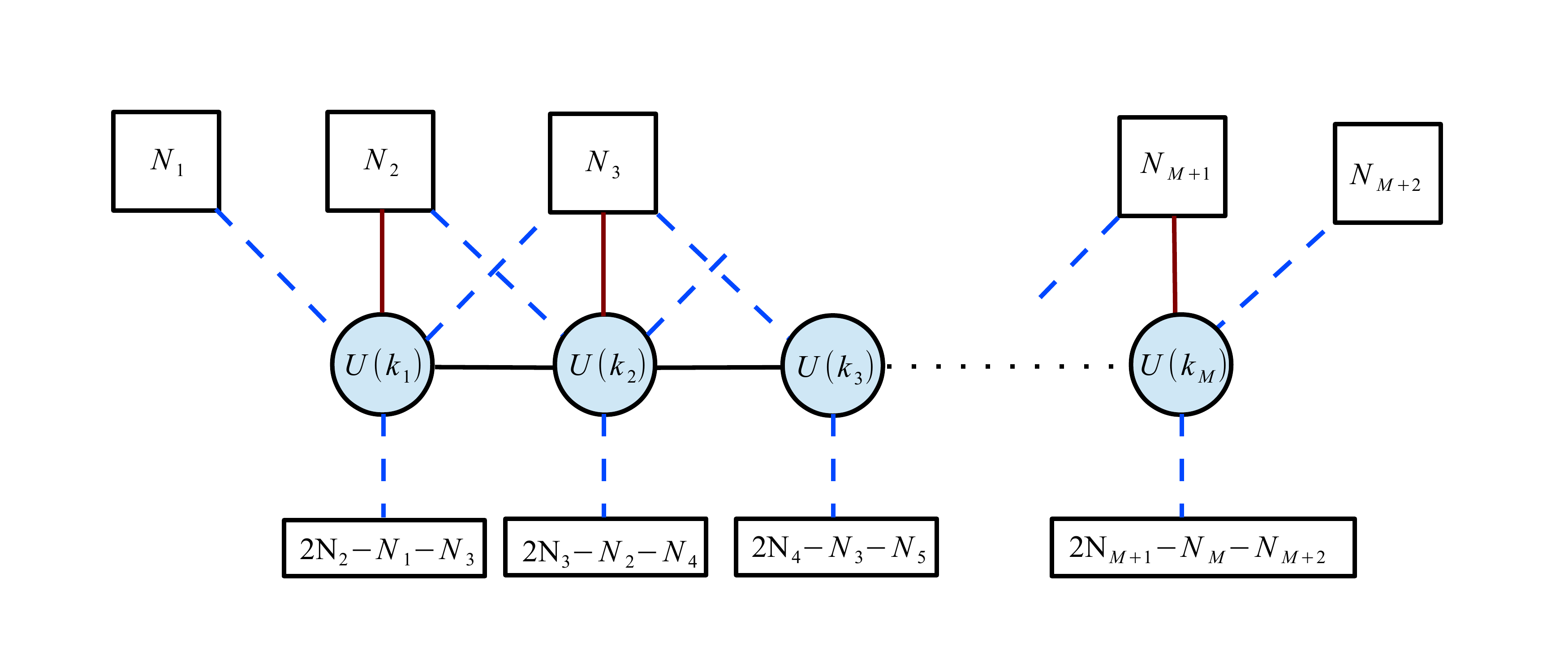}
  \caption{Convex chain of $-2$ curves. }
  \label{fig:TSU(N)convex}
\end{figure}

We can also relax the condition that all nodes should have the same $SU(N)$ flavor symmetry. In particular, we can consider the situation where the $i$th flavour node has $SU(N_i)$ symmetry together with the convexity condition
\begin{equation}
	N_i \geq \frac{1}{2}(N_{i-1} + N_{i+1}),
\end{equation}
which ensures that the parent 6d theory is not anomalous. In case of equality, all $N_i$ are ordered along a linear function and gauge anomaly cancellation is automatically satisfied. However, if $N_i$ is strictly greater than $\frac{1}{2}(N_{i-1} + N_{i+1})$ the net number of right moving fermions is greater than the number of left-moving ones and the theory will be anomalous. To cure this, we introduce for each gauge node a fourth flavor node with left-moving fermions to compensate for the excess of the  right-moving ones. The corresponding quiver is the one depicted in Figure \ref{fig:TSU(N)convex}. 
Again the elliptic genus can be computed straightforwardly using the charge table of Section \ref{sec:3}, however one has to be careful with the charge assignments of the new vertical Fermi multiplets: these are not charged under $U(1)_m, U(1)_{\epsilon_+}$ or $ U(1)_{\epsilon_-} $. The origin of the different flavor groups can be explained from the brane construction that corresponds to this theory: one has $ N_1, \dots, N_{M+2} $ D6 branes separated by NS5 branes. The difference between the number of D6 branes on the two sides of an NS5 brane must equal the negative of the cosmological constant in that region \cite{Hanany:1997sa}. So, for instance, if we have an NS5 brane with $ N_{i-1} $ D6 branes on the left and $ N_i $ on the right we must have cosmological constant $ N_{i-1}-N_{i} $ there. At the next NS5 brane, however, we must have cosmological constant $ N_{i}-N_{i+1} $. This is achieved by placing $ (N_{i}- N_{i+1}) - (N_{i-1}-N_i) = 2N_i -(N_{i-1}+N_{i+1}) $ D8 branes between the two NS5 branes \cite{Hanany:1997gh}, which has the effect of changing the cosmological constant as required. This leads to the two-dimensional quiver considered above.

\subsection{M5 branes probing a $D_{p+4}$ singularity}

A $D_{p+4}$ singularity gives rise to 7d SYM theory with gauge group $SO(2p+8)$ for $p \geq 0$. We can place $N$ M5 branes at the singularity and separate them along the remaining direction in seven dimensions. Each M5 brane actually splits into two fractional branes, which gives rise to parallel domain walls in the 7d theory \cite{DelZotto:2014hpa}. Reducing along this direction leads to a 6d $(1,0)$ SCFT with $SO(2p+8)\times SO(2p+8)$ global symmetry. Following \cite{DelZotto:2014hpa} we can obtain a Type IIA description by replacing the $D_{p+4}$ singularity with the corresponding $D_{p+4}$ ALF space and taking the circle fiber to be the M-theory circle. This results in a stack of $p+4$ parallel D6 branes on top of an $O6_{-}$ plane, together with their mirrors. Furthermore, one has $ 2N $ fractional NS5 branes \cite{Hanany:2000fq} which are of codimension 1 with respect to the orientifold plane. Whenever an $O6_{\pm}$ plane meets an NS5-brane, it turns into an $O6_{\mp}$ plane.  A system of $p+4$ D6-branes parallel to an $O6_{+}$ plane gives rise to an $Sp(p)$ gauge theory, and therefore one obtains alternating $SO(2p+8)$ and $Sp(p)$ gauge groups in 6d. On top of this, the NS5 branes contribute a total of $2N - 1$ tensor multiplets. 

\begin{figure}[h!]
  \centering
	\includegraphics[width=\textwidth]{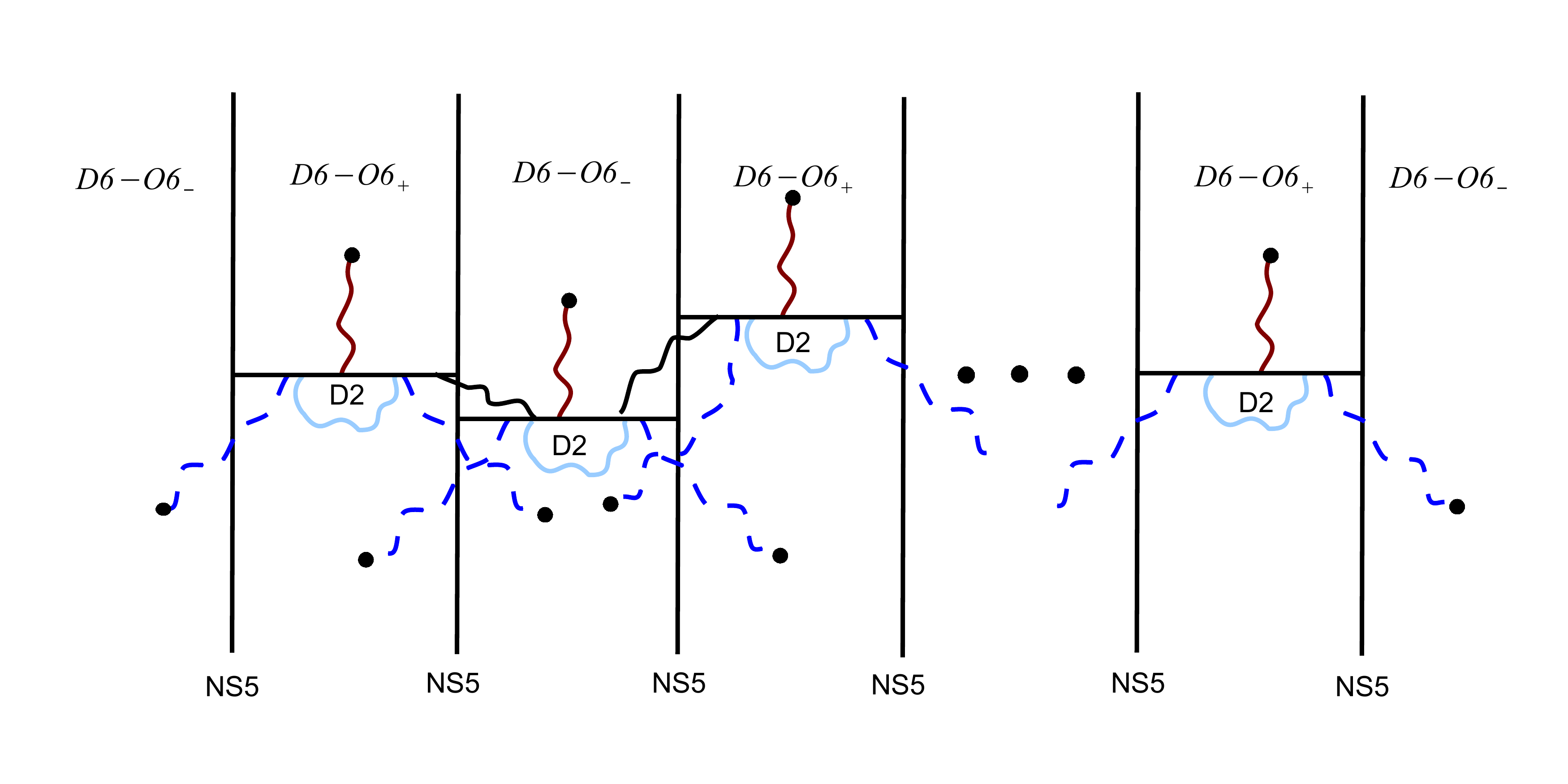}
  \caption{Type IIA brane setup corresponding to M5 branes probing $D_{p+4}$ a singularity. The fundamental strings depicted as blue or red wavy lines in this Figure give rise to fields in the 2d quiver theory. }
  \label{fig:1-4-branes}
\end{figure}

\begin{figure}[h!]
  \centering
	\includegraphics[width=\textwidth]{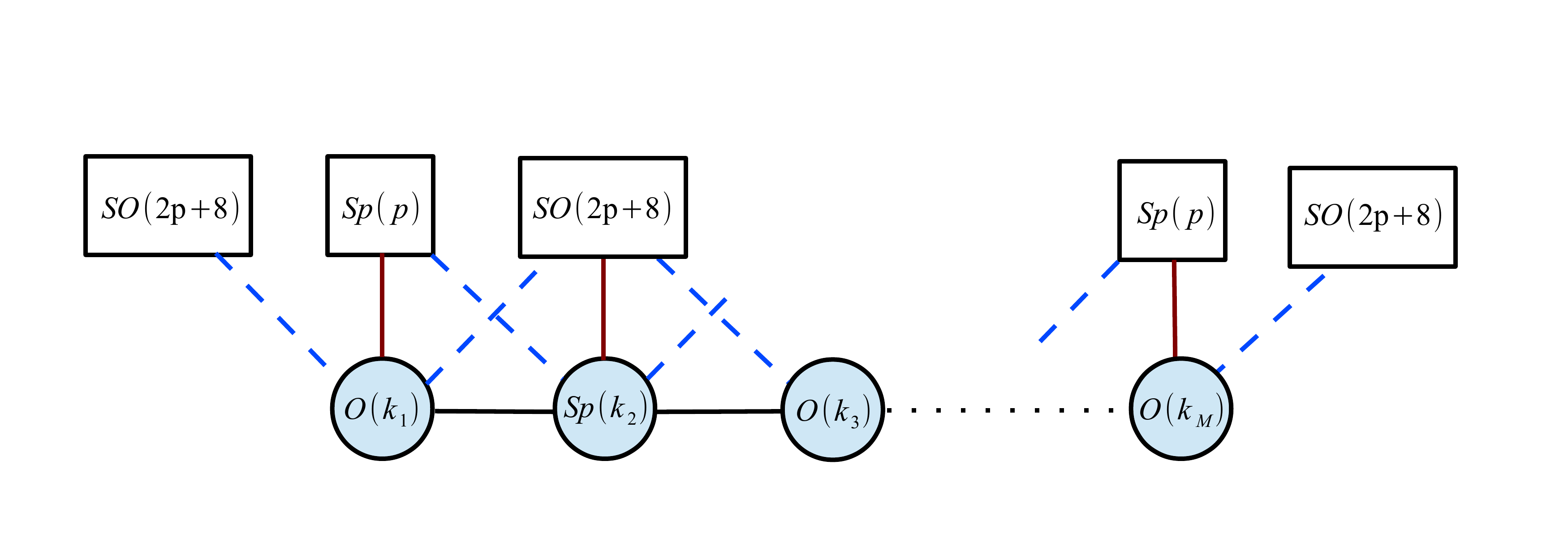}
  \caption{Non-critical strings in M5 branes probing $D_{p+4}$ singularities. }
  \label{fig:TSO}
\end{figure}

Furthermore, M2 branes suspended between M5 branes in the M-theory picture become D2 branes suspended between NS5 branes in Type IIA. The brane setup we have arrived at is pictured in Figure \ref{fig:1-4-branes}. Upon reduction along the $ X^6 $ direction, the D2 branes give rise to the self-dual strings that arise on the tensor branch of the 6d SCFT. The resulting two-dimensional quiver theory is depicted in Figure \ref{fig:TSO}. One can easily check that gauge anomalies correctly cancel out for this theory. However, one finds that $U(1)_m$ is anomalous. The reason for this anomaly can be traced back to geometry: the $D$-type singularity transverse to the M5 branes has only $SU(2)_R$ symmetry which is the $SU(2) \subset SO(4)$ which commutes with the action of the binary extension of the dihedral group. This $SU(2)_R$ is the R-symmetry group of 6d SCFT. The situation has to be contrasted with the $A$-type singularity where the surviving isometry of the space is $U(1)_m\times SU(2)_R$. Therefore, we see that $U(1)_m$ is not present for the $D$-type theory and hence the elliptic genus of it's strings should not be refined with respect to it. 

Having an explicit description of the two-dimensional quiver theory makes it possible to compute the corresponding elliptic genus. In the simplest case of a single tensor multiplet corresponding to a $ (-1) $ curve, this corresponds to the E-string elliptic genus which was studied in detail in \cite{Kim:2014dza} (although in the present setup one must identify the fugacities associated to the two $ SO(8) $ subgroups of the $ SO(16) $ flavor symmetry group). For the sake of illustration, let us also consider the non-Higgsable $ (-1)(-4)(-1) $ theory with three tensor multiplets, gauge group $ SO(8)_g $ and flavor group $ SO(8)_L\times SO(8)_R $. Let us denote by $ m_\ell^{(g)} $, $ \ell = 1,\dots,4 $ the fugacities associated to $ SO(8)_g $ and by $ m_\ell^{(L)} $ ($ m_\ell^{(R)} $)  the ones associated to $ SO(8)_L $ ($SO(8)_{R}$). From the previous discussion, one can write down the elliptic genus for any bound state of the strings associated to this theory. For instance, if one considers the bound state of one string coupled to the first $ (-1) $ tensor multiplet and one string coupled to  the $ (-4) $ multiplet, one finds:
\begin{align} \mathcal{I}_{(-1)(-4)}&=\frac{1}{4} \oint du \eta^2 \sum_{i=1}^4 \left(\frac{\eta^2}{\theta_1(\epsilon_1)\theta_1(\epsilon_2)}\right)\left(\prod_{\ell=1}^4\frac{\theta_i(m_\ell^{(L)})}{\eta}\right)\left(\prod_{\ell=1}^4\frac{\theta_i(m_\ell^{(g)})}{\eta}\right)\nonumber\\
&\times\left(\frac{\theta_1(2u)^2\theta_1(2\epsilon_+)\theta_1(2u+2\epsilon_+)\theta_1(-2u+2\epsilon_+)}{\eta^3\theta_1(\epsilon_1)\theta_1(\epsilon_2)}\right)\nonumber\\
& \times \left(\prod_{\ell=1}^4\frac{\eta^4}{\theta_1(\epsilon_+\!+\!m^{(g)}_\ell\!+\!u)\theta_1(\epsilon_+\!+\!m^{(g)}_\ell\!-\!u)\theta_1(\epsilon_+\!-\!m^{(g)}_\ell\!+\!u)\theta_1(\epsilon_+\!-\!m^{(g)}_\ell\!-\!u)}\right)\nonumber\\
&\times\left(\frac{\theta_i(\epsilon_- - u)\theta_i(\epsilon_- + u)}{\theta_i(-\epsilon_+- u)\theta_i(-\epsilon_+ +u)}\right),\end{align}
where 
\begin{equation} \eta = q^{1/24}\prod_{i=1}^\infty(1-q^i),\qquad q = e^{2\pi i \tau}\end{equation}
and
\begin{equation}\theta_1(x) = i \, q^{1/8}e^{-\pi i x}(1-e^{2\pi i x})\prod_{i=1}^\infty (1-q^i)(1-q^i e^{2\pi i x})(1-q^i e^{-2\pi i x}) .\end{equation}
The contour integral can be performed by using the Jeffrey-Kirwan prescription for computing residues \cite{Benini:2013xpa}. Similarly, one can compute the elliptic genus for other bound states of strings.

\section{Multiple M5 branes probing an M9 wall}

\begin{figure}[h!]
 \centering
	\includegraphics[width=0.8\textwidth]{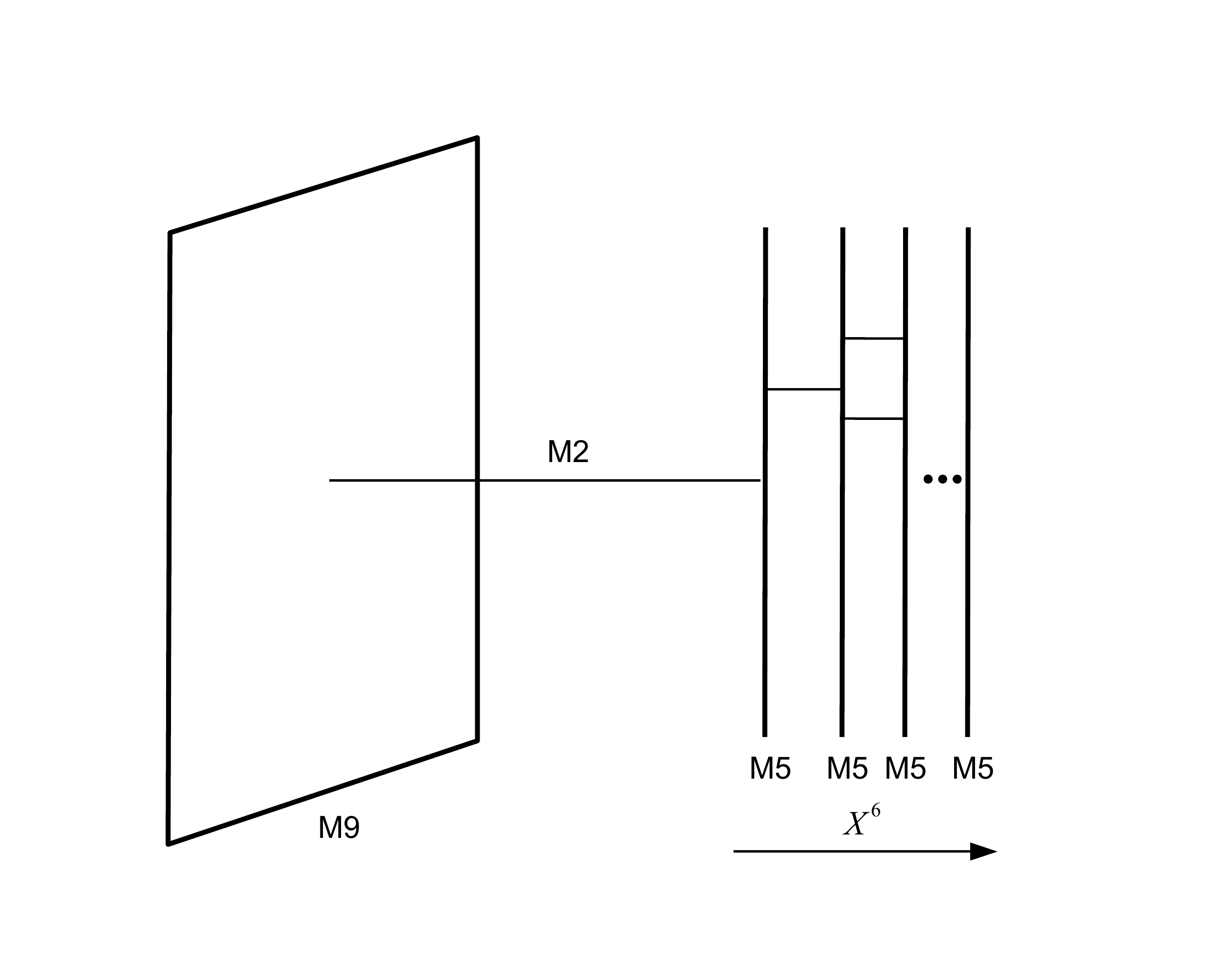}
 \caption{E-strings as suspended M2 branes between M5 branes probing an M9 wall.}
 \label{fig:M9-M5}
\end{figure}

In this section we study the $ \mathcal{N} = (1,0) $ six-dimensional theory of $ N $ small $ E_8 $ instantons \cite{Ganor:1996mu,Seiberg:1996vs}; upon moving to the tensor branch, this becomes the theory of $ N $ parallel M5 branes in the proximity of the M9 boundary wall of M-theory. The strings originate from M2 branes that are suspended between neighboring M5 branes or between the M5 branes and the M9 plane (see Figure \ref{fig:M9-M5}). Upon circle reduction to five dimensions with an $E_8$ background Wilson line 
(which breaks $E_8$ global symmetry to $SO(16)$), the theory of $ N $ small instantons reduces to 
the  $Sp(N)$ theory with $8$ fundamental and $1$ antisymmetric hypermultiplets \cite{Seiberg:1996vs}. The instanton calculus for this five-dimensional theory provides a way to check elliptic genus computations and will be exploited in Section \ref{sec:Sp2calculus}.\\

\subsection{Two-dimensional quiver}

In order to derive a quiver description for the theory of the strings, it again proves useful to switch to an equivalent brane configuration within string theory. Let us begin by discussing $ N = 1 $ theory of a single small $ E_8 $ instanton, whose associated two-dimensional quiver gauge theory has been worked out in \cite{Kim:2014dza}. The quiver was derived from a Type I' brane configuration, which arises as follows: upon reduction of M-theory on a circle, the M9 plane is replaced by eight D8 branes on top of an O$ 8^- $ orientifold plane (which has D8 brane charge $ -8 $); the M5 brane, on the other hand, becomes an NS5 brane. Furthermore, M2 branes are replaced by D2 branes that stretch between the NS5 brane and the D8-O8$^-$ system. By studying the two-dimensional reduction of the worldvolume theory of the D2 branes in the limit of small separation between the NS5 and eight-branes, one arrives at the two-dimensional quiver gauge theory of \cite{Kim:2014dza}. The (0,4) quiver gauge theory for $ n $ strings has gauge group $ O(n) $ and the following field content: a vector multiplet in the adjoint (anti-symmetric) representation of $ O(n) $, a hypermultiplet in the symmetric representation, and eight Fermi multiplets in the bifundamental representation of $ O(n)\times SO(16) $. Elliptic genera for this theory have been computed in \cite{Kim:2014dza} for up to four strings and shown to agree with results from the instanton calculus for the five-dimensional $ Sp(1) $ theory with eight fundamental hypermultiplets.

\begin{figure}[h!]
 \centering
	\includegraphics[width=5.5in]{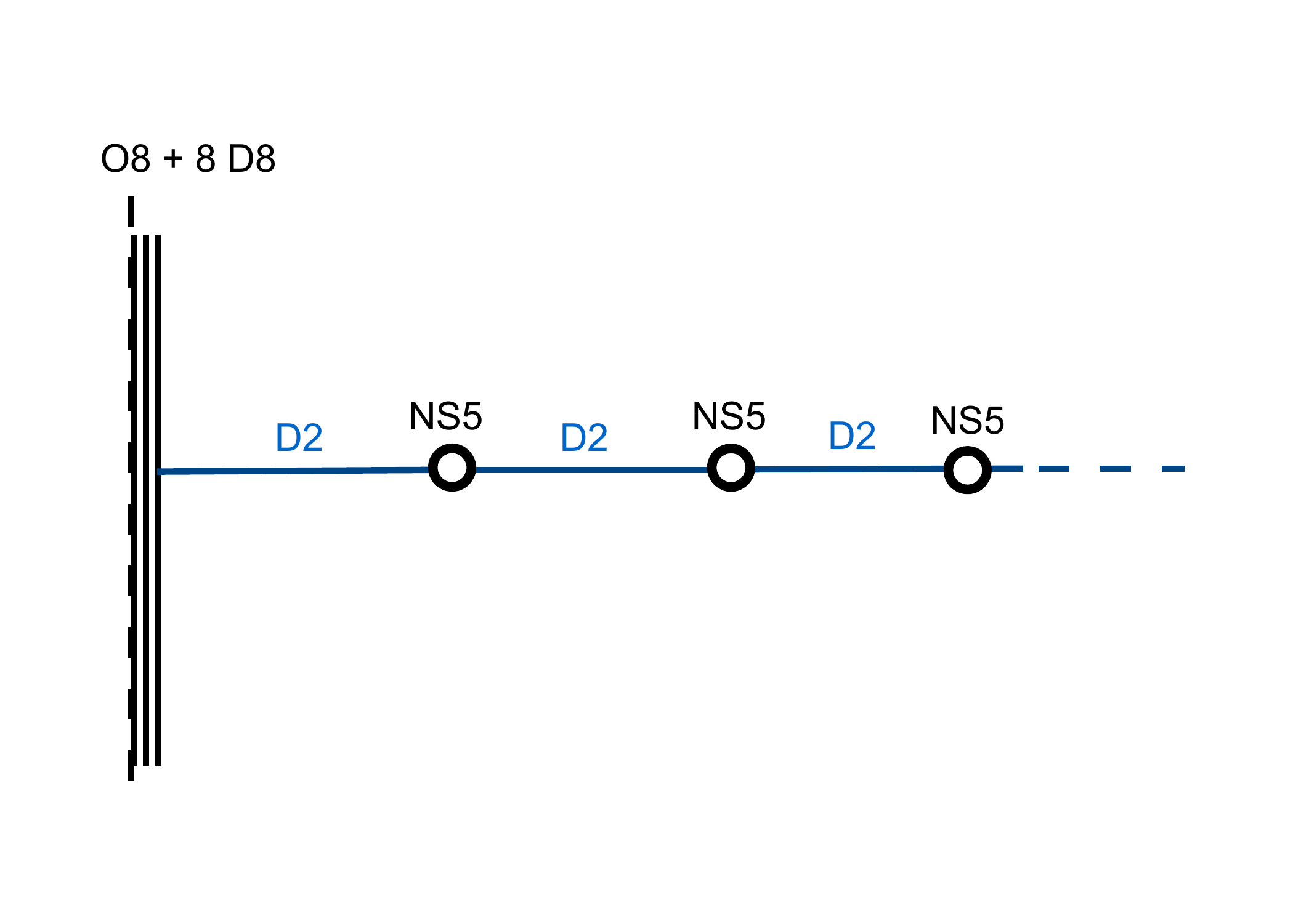}
 \caption{Brane configuration for the theory of $N$ small $ E_8 $ instantons.}
 \label{fig:1222chain}
\end{figure}

The generalization of the Type I' brane setup to the case of $ N $ small instantons is illustrated in Figure \ref{fig:1222chain} and is again obtained by reducing the above M9-M5 setup on a circle. The brane setup is a combination of the E-string and M-string brane configurations without the D6 branes which are usually present for the M-string system. As we will see this becomes crucial when we look at the quiver-gauge theory governing the dynamics of the strings to which we now turn.

\begin{figure}[h!]
 \centering
	\includegraphics[width=5in]{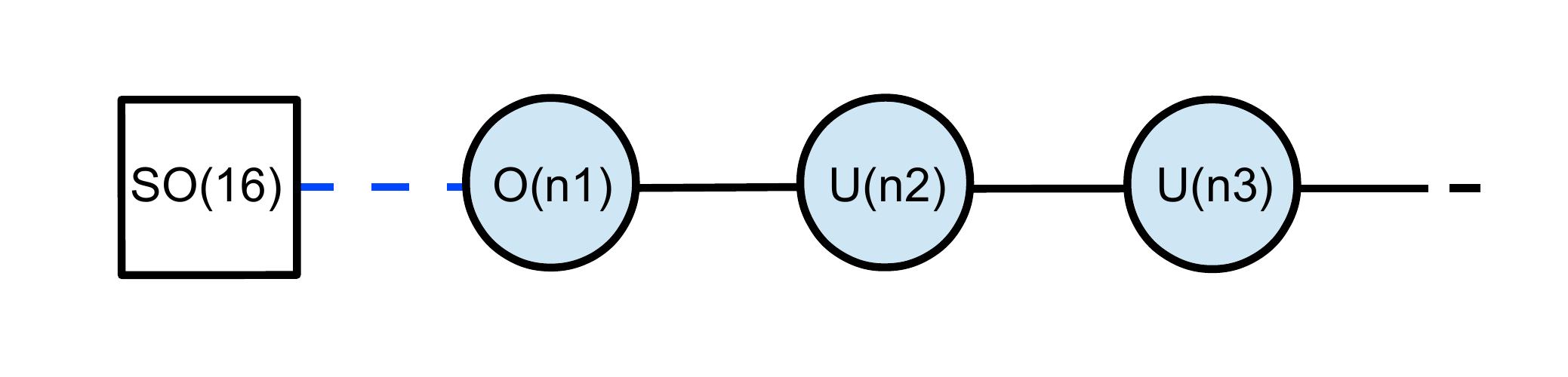}
 \caption{Quiver for the theory of $N$ small $ E_8 $ instantons.}
 \label{fig:1222quiver}
\end{figure}

The brane setup implies a simple quiver gauge theory governing the dynamics of the strings. The first $ n_1 $ D2 branes ending on the D8-O8 system correspond to a $ O(n_1) $ gauge node; from the D2-D8  strings one finds eight bifundamental Fermi multiplets charged under $ O(n_1)\times SO(16) $. Furthermore, there is a symmetric hyper at the $O(n_1)$ node as already observed in \cite{Kim:2014dza}. All other gauge nodes corresponding to the the D2 branes suspended between NS5 branes have unitary gauge groups with bi-fundamental matter between them familiar from the orbifolds of M-strings \cite{Haghighat:2013tka}. Finally, one also obtains $ O(n_1)\times U(n_2) $ bifundamentals from strings ending on the $ n_1 $ and $ n_2 $ D2 branes. These bifundamental fields consist of a (0,4) hyper and a (0,4) Fermi multiplet, as is the case for M-strings. The resulting quiver is illustrated in Figure \ref{fig:1222quiver}. 

We comment on the global symmetries of this quiver gauge theory, and compare them with 
the symmetries that we expect for the infrared CFT on these strings. Our $(0,4)$ gauge theory 
has $SO(4)=SU(2)\times SU(2)$ R-symmetry. The first $SU(2)$ is part of the $SO(4)$ symmetry 
which rotates $\mathbb{R}^4$ along the worldvolume of NS5-branes, transverse to the strings. 
The second $SU(2)\sim SO(3)$ rotates the $\mathbb{R}^3$ space transverse to the NS5-branes 
and D2-branes. The infrared (or equivalently strong coupling) limit of the 2d gauge theory
is realized by going to the M-theory regime of the type I' theory. Then the space $\mathbb{R}^3$ 
transverse to NS5-D2 is replaced by $\mathbb{R}^3\times S^1$, including the M-theory circle, 
and becomes $\mathbb{R}^4$ in the strong coupling limit. So in the IR, we expect the $SO(3)$ 
symmetry to enhance to $SO(4)$. Any analysis from our gauge theory, such as the elliptic genus 
calculus below, will be missing the extra Cartan charges of the enhanced $SO(4)$. Let us denote
by $\epsilon_{1,2}$ the chemical potentials for the rotations on $\mathbb{R}^4$, 
as in the previous sections. Apart from rotating $\mathbb{R}^4$ along the 5-brane, there will be 
an extra rotation on $\mathbb{R}^4$ transverse to the M5-brane, with 
$\epsilon_+\equiv\frac{\epsilon_1+\epsilon_2}{2}$. Let us denote by $m$ the chemical potential 
for the missing Cartan of the enhanced IR symmetry. Then the $\mathbb{R}^3$ part in the type I' 
setting is rotated by $m+\epsilon_+$, while the rotation by $m-\epsilon_+$ is invisible 
on $\mathbb{R}^3\times S^1$. Thus, our UV gauge theory will be computing the elliptic genus 
only at $m=\epsilon_+$\footnote{The parameter $m$ is the one appearing in the instanton calculus of the Nekrasov partition function and should not be confused with the fugacity of $U(1)_m$ in the 2d gauge theory. With respect to the 2d fugacity it is shifted by $\epsilon_+$.}. 

At $N=1$, it is known that the 6d SCFT engineered by a single M5 and M9 brane does not see the 
extra Cartan of $SO(4)$ (conjugate to $m$) at all. In other words, all the states 
in the Hilbert space of the 6d SCFT are completely neutral under this $U(1)$ charge (while the full M-theory 
would see the charged states decoupled from the 6d SCFT). One way to see this is from the 
5 dimensional $Sp(N)$ gauge theory obtained after circle compactification. Namely, the parameter 
$m$ above corresponds to the mass parameter for the $Sp(N)$ antisymmetric hypermultiplet in the resulting 
5d theory. At $N=1$, the antisymmetric representation is neutral in $Sp(1)$ and the corresponding 
hypermultiplet decouples. This implies that the extra $U(1)$ for $m$ decouples from the 6d CFT 
at $N=1$, and this has been tested from the instanton partition function in \cite{Hwang:2014uwa}.
This is the reason why the 2d gauge theory above provided maximally refined elliptic genera at $N=1$ in \cite{Kim:2014dza}, since the restriction $m=\epsilon_+$ loses no information about the 6d SCFT.
However, the parameter $m$ appears in the 6d CFT spectrum for $N\geq 2$, 
which was checked from the $Sp(N)$ instanton calculus \cite{Hwang:2014uwa}.

Below we present sample computations for the elliptic genera corresponding to the lowest charge sectors, namely $(n_1,n_2) = (1,1)$, $(n_1,n_2) = (1,2)$, and $(n_1,n_2) = (2,1)$ for the $N=2$ quiver.

\subsubsection*{Charge sector $(n_1, n_2) = (1,1)$}

Combining the one-loop determinants, the zero-mode integral $I_{(1,1)}$ is given by
\begin{align}
    -\oint \frac{du}{2} \sum_{i=1}^{4} \Bigg(\frac{\eta^2}{\theta_1( \epsilon_1)\theta_1( \epsilon_2)}\Bigg)_1 \!\Bigg(\frac{\eta^3 \theta_1 (\epsilon_1 \!+\! \epsilon_2)}{\theta_1( \epsilon_1)\theta_1( \epsilon_2)}\Bigg)_2 \! \Bigg(\prod_{l=1}^{8}\frac{\theta_1 (m_l\!+\!a_i)}{\eta}\Bigg)_3 \! \Bigg( \frac{\theta_1 (+\epsilon_- \!\pm\! (a_i \!- \! u))}{\theta_1 (-\epsilon_+ \!\pm\! (a_i \!-\! u))}\Bigg)_{4}
\end{align}
where $a_i = (0, \frac{1}{2}, \frac{1+\tau}{2}, \frac{\tau}{2})_i$. Repeated signs $\pm$ in the 
arguments mean that both factors are multiplied: $\theta_1(-\epsilon_+ \pm (a_i - u)) 
\equiv \theta_1(-\epsilon_+ + a_i - u) \theta_1(-\epsilon_+ - a_i + u)$.
The contour integral given by the JK-Res is done with $\eta = e_1$. Then the only nonzero JK-Res 
comes from the pole $u = a_i +\epsilon_+$. The result is
\begin{align}
    I_{(1,1)} =& -\frac{\eta^5 \theta_1 (\epsilon_1 + \epsilon_2)}{2\theta_1(\epsilon_1)^2 \theta_1(\epsilon_2)^2} \sum_{i=1}^{4}  \Bigg(\prod_{l=1}^{8}\frac{\theta_1 (m_l+a_i)}{\eta}\Bigg) \frac{\theta_1 (\epsilon_1) \theta_1 (-\epsilon_2)}{\eta^3 \theta_1(-\epsilon_1 - \epsilon_2)}\\
    =&-\frac{\eta^2}{2\theta_1(\epsilon_1) \theta_1 (\epsilon_2)} \sum_{i=1}^{4}  \Bigg(\prod_{l=1}^{8}\frac{\theta_i (m_l)}{\eta}\Bigg)=I_{(1,0)}\ .\nonumber
\end{align}
This is the elliptic genus of the single E-string, 
i.e. with charge $(n_1,n_2)=(1,0)$ \cite{Klemm:1996hh,Kim:2014dza}.

\subsubsection*{Charge sector $(n_1, n_2) = (1,2)$}

The zero-mode integral is given by
\begin{align}
    I_{(1,2)} =& -\oint \frac{du_1 du_2}{4} \; \sum_{i=1}^{4} \frac{\eta^2}{\theta_1( \epsilon_1)\theta_1( \epsilon_2)} \Bigg(\frac{\eta^3 \theta_1 (\epsilon_1 + \epsilon_2)}{\theta_1( \epsilon_1)\theta_1( \epsilon_2)}\Bigg)^{2}  \Bigg(\prod_{l=1}^{8}\frac{\theta_1(m_l+a_i)}{ \eta}\Bigg) \nonumber \\
    &\times \frac{\theta_1 ( \pm(u_1\!-\!u_2)) \theta_1 ( \epsilon_1 \!+\! \epsilon_2 \!\pm\!(u_1\!-\!u_2))}{\theta_1 ( \epsilon_1 \!\pm\!(u_1\!-\!u_2)) \theta_1 ( \epsilon_2 \!\pm\!(u_1\!-\!u_2))}
    \frac{\theta_1(+\epsilon_- \!\pm\! a_i \!\mp\! u_{1})}{\theta_1(-\epsilon_+ \!\pm\! (a_i\!-\! u_{1}))} \frac{\theta_1(+\epsilon_- \!\pm\! a_i \!\mp\! u_{2})}{\theta_1(-\epsilon_+ \!\pm\! a_i \!\mp\! u_{2})}.
\end{align}
If we choose $\eta = e_1 + \epsilon \, e_2$ in which $\epsilon \ll 1$, nonzero JK-Res can 
only come from the following poles.
\begin{itemize}
	\item $(\epsilon_{1,2}-u_1+u_2 , -\epsilon_+ - a_i + u_1) = (0,0)$
	\item $(\epsilon_{1,2}+u_1-u_2 , -\epsilon_+ - a_i + u_2) = (0,0)$
	\item $(-\epsilon_+ - a_i + u_1 , -\epsilon_+ - a_i + u_2) = (0,0)$.
\end{itemize}
Actually evaluating the residues, it turns out that all these poles yield 
vanishing residues, so that
\begin{align}
    I_{(1,2)} =& 0.
\end{align}

\subsubsection*{Charge sector $(n_1,n_2) = (2,1)$}

Now the gauge theory comes with $O(2)\times U(1)$ gauge group. In the elliptic genus calculus, 
there are seven disconnected sectors of $O(2)$ flat connections \cite{Kim:2014dza}. In six sectors, 
the flat connections are discrete, while in one sector it comes with one complex parameter.

In the first sector with continuous parameter, which we label by superscript $0$, 
one has to do the following rank $2$ contour integral for the elliptic genus:
\begin{align}\label{(2,1)-continuous}
   I_{(2,1)}^{0} =&  \oint \frac{du_1 du_2}{2} \; \frac{\eta^7 \theta_1 (\epsilon_1 + \epsilon_2)}{\theta_1( \epsilon_{1,2}) \theta_1 (\epsilon_{1,2} \pm 2 u_1)} \frac{\eta^3 \theta_1 (\epsilon_1 + \epsilon_2)}{\theta_1( \epsilon_1)\theta_1( \epsilon_2)}  \Bigg(\prod_{l=1}^{8}\frac{\theta_1(m_l \pm u_1)}{ \eta^2}\Bigg) \\
   &\times \frac{\theta_1(+\epsilon_- \pm u_1 + u_2)}{\theta_1(-\epsilon_+ \pm u_1 + u_2)} \frac{\theta_1(+\epsilon_- \pm u_1 - u_2)}{\theta_1(-\epsilon_+ \pm u_1 - u_2)} \nonumber
\end{align}
If we choose $\eta = e_1 + \epsilon \, e_2$ in which $\epsilon \ll 1$, nonzero JK-res 
can only appear from the following poles.
\begin{itemize}
	\item $(-\epsilon_+ - u_1 + u_2, \epsilon_{1,2} + 2 u_1) = (0,0) \longrightarrow (u_1, u_2) = (-\tfrac{\epsilon_{1,2}}{2} + a_i, \tfrac{\epsilon_{2,1}}{2} + a_i)$. Its residue is zero.
	\item $(-\epsilon_+ + u_1 - u_2, -\epsilon_+ + u_1 + u_2) = (0,0) \longrightarrow (u_1, u_2) = (\epsilon_+ + a_i, a_i)$
	\item $(-\epsilon_+ + u_1 + u_2, \epsilon_{1,2} + 2 u_1) = (0,0) \longrightarrow (u_1, u_2) = (-\tfrac{\epsilon_{1,2}}{2} + a_i, \epsilon_{1,2} + \tfrac{\epsilon_{2,1}}{2} - a_i)$
\end{itemize}
Collecting all the residues, $I_{(2,1)}^{0}$ is given by
\begin{align}
I_{(2,1)}^{0} &= \frac{\eta^{-12}}{4 \theta_1 (\epsilon_1) \theta_1 (\epsilon_2)} \sum_{i=1}^{4} \Bigg[\frac{\prod_{l=1}^{8} \theta_i (m_l \pm \epsilon_+)}{\theta_1 (2 \epsilon_1 + \epsilon_2) \theta_1 (\epsilon_1 + 2 \epsilon_2)}\nonumber\\
&\hspace{1.5in}-\Bigg(\frac{ \theta_1 (\epsilon_1 + \epsilon_2) \prod_{l=1}^{8} \theta_i (m_l \pm \tfrac{\epsilon_1}{2} )}{\theta_1(\epsilon_1 ) \theta_1 ( \epsilon_1 - \epsilon_2) \theta_1 (2\epsilon_1 +  \epsilon_2)} + (\epsilon_1 \leftrightarrow \epsilon_2)\Bigg) \Bigg]\nonumber\\
&= \frac{\eta^{4}}{4 \theta_1 (\epsilon_1)^2 \theta_1 (\epsilon_2)^2} \sum_{i=1}^{4} \Bigg(\prod_{l=1}^{8} \frac{\theta_i (m_l)^2}{\eta^2}\Bigg)\ .
\end{align}
On the second line, we used the following identity 
\begin{align*}
    \sum_{i=1}^{4} &\prod_{l=1}^{8} \theta_i(m_l)^2 = \sum_{i=1}^{4}\,\,
    \!\!\Bigg[\frac{\theta_1(\epsilon_{1,2}) \prod_{l=1}^{8} \theta_i(m_l \pm \epsilon_+)}{\theta_1(2 \epsilon_1 + \epsilon_2) \theta_1 (\epsilon_1 + 2 \epsilon_2)}\nonumber\\
    &\hspace{1.35in}- \Bigg(\frac{\theta_1(\epsilon_{2}) \theta_1(\epsilon_1 + \epsilon_2) \prod_{l=1}^{8} 
    \theta_i(m_l \pm \frac{\epsilon_1}{2})}{\theta_1(2 \epsilon_1 + \epsilon_2) \theta_1 (\epsilon_1 - \epsilon_2)} + (\epsilon_1 \leftrightarrow \epsilon_2) \Bigg)\Bigg],
\end{align*}
which we checked in an expansion in $e^{2\pi i\tau}$, 
for the first $5$ terms up to $(e^{2 \pi i \tau})^{5/2}$ order.

The contributions from the other six sectors are given by
\begin{align}
    I_{(2,1)}^{m} &= \oint \frac{d u}{4} \frac{\eta^4 \theta_1 (a_1 + a_2) \theta_1 (\epsilon_1 + \epsilon_2 + a_1 + a_2) }{\theta_1 (\epsilon_{1,2})^2 \theta_1(\epsilon_{1,2} + a_1 + a_2)} \frac{\eta^3 \theta_1 (\epsilon_1 + \epsilon_2)}{\theta_1( \epsilon_1)\theta_1( \epsilon_2)} \nonumber\\
    &\times \Bigg( \prod_{l=1}^{8} \frac{\theta_1 (m_l + a_1)\theta_1 (m_l + a_2)}{\eta^2}\Bigg)  \frac{\theta_1 (+\epsilon_- \pm a_1 \mp u )}{\theta_1 (-\epsilon_+ \pm a_1 \mp u)} \frac{\theta_1 (+\epsilon_- \pm a_2 \mp u )}{\theta_1 (-\epsilon_+ \pm a_2 \mp u)},
\end{align}
where we take the discrete $O(2)$ holonomies  $(a_1, a_2) = (0, \frac{1}{2})$, $(\frac{\tau}{2}, \frac{1+\tau}{2})$, $(0, \frac{\tau}{2})$, $(\frac{1}{2}, \frac{1+\tau}{2})$, $(0, \frac{1+\tau}{2})$,
$(\frac{1}{2}, \frac{\tau}{2})$ for $m = 1, 2,\  \cdots, 6$, respectively. JK-res with 
$\eta = e_1$ can be nonzero only at the pole $u = a_1 +\epsilon_+$ or $u=a_2+\epsilon_+$, 
yielding the following result:
\begin{align}\label{(2,1)-others}
    I_{(2,1)}^{m} &= \frac{\eta^4  }{2\theta_1 (\epsilon_{1,2})^2 }   \prod_{l=1}^{8} \frac{\theta_i (m_l)\theta_j (m_l)}{\eta^2}
\end{align}
where $(i,j) = (1,2),(4,3),(1,4),(2,3),(1,3),(2,4)$ for $m = 1, 2,\  \cdots, 6$. 
Combining (\ref{(2,1)-continuous}) and (\ref{(2,1)-others}), one obtains
\begin{align}
    I_{(2,1)} = I_{(2,1)}^{0} + \sum_{m=1}^{6} I_{(2,1)}^{m} = \left(I_{(1,1)}\right)^2
    =\left(I_{(1,0)}\right)^2\ ,
\end{align}
which exhibits a factorization structure.

\vskip .5cm

In the next subsection, we will show that all the results above are in complete agreement with 
the 5 dimensional $Sp(2)$ instanton calculus of \cite{Hwang:2014uwa}. Before that, let us first 
try to interpret these rather simple results that we have found at $m=\epsilon_+$.

The strings made of $n_1$ and $n_2$ D2-branes in Fig. 6, winding a circle, contribute to 
the elliptic genus as both $n_1+n_2$ multi-particle states, and also through various threshold bound 
states with lower particle numbers.
There could be various kinds of bound states. Generally, $m_1(\leq n_1)$ of the $n_1$ strings 
and $m_2(\leq n_2)$ of the $n_2$ strings may form bounds. One can first deduce that the index 
is zero at $m=\epsilon_+$ in the sector which contains bound states with charges $(0,m_2)$. 
This is because the $(0,m_2)$ bounds are basically M-strings in a maximally supersymmetric theory.
Note that the M-strings are half-BPS states of the 6d $(2,0)$ theory,
so will see $8$ broken supercharges as Goldstone fermions. This is in contrast to the
strings in 6d QFTs preserving $(1,0)$ SUSY only. The extra fermionic zero modes for M-strings
provide the factor
\begin{equation}
  \sin\pi(m+\epsilon_+)\sin\pi(m-\epsilon_+)
\end{equation}
to the elliptic genus \cite{Kim:2011mv}. Thus, M-strings which are unbound to E-strings
(i.e. at $m_1=0$) will contribute a $0$ factor to the elliptic genus at $m=\pm\epsilon_\pm$.

With this understood, let us start by considering the sector with $(n_1,n_2)=(1,1)$. 
At $m=\epsilon_+$, there is no contribution from the two particle states $(1,0)+(0,1)$
due to the above reasoning. So one should only obtain a single particle contribution in 
the $(1,1)$ sector. This is consistent with our finding $I_{(1,1)}=I_{(1,0)}$. A slightly 
surprising fact from our finding is that the single particle bound with charges $(1,1)$ 
behaves exactly the same as a single E-string with charge $(1,0)$, at least at $m=\epsilon_+$. 
Although the $(1,1)$ bound look like one long E-string suspended between the M9-plane and 
the second M5-brane, it penetrates through the first M5-brane so in principle there could 
be extra contributions to the BPS degeneracies from the intersection. For instance, 
in the case of M-strings, it is known that the charge $(1,0)$ M-string and the single particle 
bound part of the $(1,1)$ M-string exhibit different spectra (at general
chemical potentials, with $m\neq \epsilon_+$) \cite{Kim:2011mv}. So we interpret that 
$I_{(1,1)}=I_{(1,0)}$ implies some simplification of the $(1,1)$ elliptic genus at 
$m=\epsilon_+$.

Other results also have nontrivial implications on the E-/M-string bound state elliptic genera 
at $m=\epsilon_+$. For $(n_1,n_2)=(1,2)$, $I_{(1,2)}=0$ implies that there are no $(1,2)$ 
bound states captured by the elliptic genus at $m=\epsilon_+$, since we know that 
$(1,1)+(0,1)$, $(1,0)+(0,2)$, or $(1,0)+2(0,1)$ multi-particles cannot contribute to 
the elliptic genus at $m=\epsilon_+$. As we will consider from the 5d $Sp(2)$ instanton 
calculus, this feature generalizes to higher string numbers: the $(m_1,m_2)$ bounds with 
$m_1<m_2$ do not contribute to the elliptic genus at $m=\epsilon_+$. 

Finally, $I_{(2,1)}=I_{(1,0)}^2$ can also be understood with the above observations.
Namely, with a $(0,m_2)$ particle yielding a factor of zero in the elliptic genus, 
the nonzero contribution can come from $(1,1)+(1,0)$ two particle states. But since 
we already know that these contributions give equal elliptic genera namely that of a single E-string, we 
can naturally understand this relation. (So our finding implies that the $(2,1)$ bound does 
not contribute to the index at $m=\epsilon_+$.)

Based on the above observations, we propose that
\begin{eqnarray}\label{factorization}
  I_{(n_1,n_2)}&=&0\ \ {\rm if }\ \ n_1<n_2\\
  &=&I_{(n_2,0)}I_{(n_1-n_2,0)}\ \ {\rm if}\ \ n_1\geq n_2\ .\nonumber
\end{eqnarray}
Namely, at $m=\epsilon_+$, the $(n_1,n_2)$ string elliptic genus factorizes 
to two E-string elliptic genera. Although we have shown this result for only a 
few charges from the 2d gauge theories, we shall confirm such factorizations to 
a much higher order in $n_1,n_2$ from the 5d $Sp(2)$ instanton calculus below.

\subsection{Five dimensional $Sp(2)$ instanton calculus}\label{sec:Sp2calculus}

In this subsection, we shall consider the circle compactification of the $(1,0)$ theory on 
$2$ M5 and one M9, and consider the string spectra from the instanton calculus of the resulting 5d gauge theory.

Let us consider the six-dimensional conformal field theory living on two M5 branes 
probing the M9 plane. The space transverse to the two M5-branes is $\mathbb{R}^4\times\mathbb{R}^+$, 
where the latter $\mathbb{R}^+$ is obtained by the $\mathbb{Z}_2$ action of M9. This space 
has $SO(4)=SU(2)\times SU(2)$ symmetry. The first $SU(2)$ is the superconformal R-symmetry, 
and the second $SU(2)$ is a flavor symmetry. The full flavor symmetry is thus $SU(2)\times E_8$.

We compactify this system on a circle, with an $E_8$ Wilson line which breaks $E_8$ into $SO(16)$.
Then at low energy, one obtains a 5 dimensional $\mathcal{N} = 1$ $Sp(N)$ gauge theory with 
$N_f = 8$ fundamental and one anti-symmetric hypermultiplet. The $8$ masses $\tilde{m}_i$ for 
the fundamental hypermultiplets and the mass $m$ for the antisymmetric hypermultiplet are in 1-1 
correspondence to the chemical potentials of the $E_8\times SU(2)$ flavor symmetries.
The precise relations that we use are given in \cite{Hwang:2014uwa,Kim:2014dza}. $m$ is simply 
uplifting to the $SU(2)$ flavor chemical potential, while the masses $\tilde{m}_i$ are related 
to the $E_8$ chemical potentials $m_i$ by \cite{Kim:2014dza}
\begin{equation}
  \tilde{m}_i=m_i\ \ \ ({\rm for}\ \ i=1,\cdots,7)\ ,\ \ 
  \tilde{m}_8=m_8-\tau\ .
\end{equation}
The chemical potentials $\tilde{\alpha}_1$, $\tilde{\alpha}_2$ 
for the $Sp(2)$ electric charges are related to those
$\alpha_{1,2}$ for the string winding numbers $n_1-n_2,n_2$ by
\begin{equation}
  \tilde\alpha_1=\alpha_1+\frac{\tau}{2}-m_8\ ,\ \ \tilde\alpha_2=\alpha_2+\frac{\tau}{2}-m_8\ .
\end{equation}
We chose the convention for $\alpha_1,\alpha_2$ in a way that they correspond to 
the distances from the M5-branes to the M9-plane. 

The elliptic genera of the previous subsection are related to the instanton 
partition function for this 5d $Sp(2)$ Yang-Mills, as follows.
Let us first define the fugacities
\begin{align}
  s = e^{2\pi i\epsilon_+}, u = e^{2\pi i\epsilon_-}, v = e^{2\pi im}, 
  \tilde{w}_i = e^{2\pi i\tilde\alpha_i}, \tilde{y}_i = e^{2\pi i\tilde{m}_i}
\end{align}
where $\tilde{w}_i$ satisfy $\tilde{w}_2<\tilde{w}_1<1$ to probe the sectors 
with $n_1>0$ and $n_2>0$. One should first consider the perturbative partition function 
$Z^{Sp(2)}_\text{pert}$, given by
\begin{align}
  Z^{Sp(2)}_\text{pert} = PE\Bigg[ &-\frac{s(s + s^{-1})}{(1-s u)(s u^{-1})} 
  \Big[ \tilde{w}_1 \tilde{w}_2 + \tilde{w}_2 / \tilde{w}_1 + \tilde{w}_2^{2} + \tilde{w}_1^{2} 
  \Big]\nonumber\\ & \hspace{.78in}+ \frac{s(v + v^{-1})}{(1-s u)(1-s u^{-1})} 
  \Big[ \tilde{w}_1 \tilde{w}_2 + \tilde{w}_2 / \tilde{w}_1\Big]\nonumber\\
  &\hspace{1in}+\sum_{i=1}^8\frac{s(\tilde y_i + \tilde y_i^{-1})}{(1-s u)(1-s u^{-1})} 
  \Big[ \tilde{w}_1 + \tilde{w}_2 \Big]\Bigg],
\end{align}
where $\text{PE}[f]$ is defined as
\begin{align}
    \text{PE}[f(s,u,v,\tilde{w}_{1,2},\tilde{y}_i,q)] = \exp\left[\sum_{n=1}^{\infty} \frac{1}{n}f(s^n,u^n,v^n,\tilde{w}_{1,2}^n,\tilde{y}_i^n,q^n)\right].
\end{align}
The instanton part $Z^{Sp(2)}_\text{inst}$ is computed from the ADHM quantum mechanics. 
It is well known (see, for instance \cite{Hwang:2014uwa} and references therein) that the ADHM 
calculus sometimes captures extra decoupled contributions apart from the field theory index.
This decoupled contribution is computed in our case in \cite{Hwang:2014uwa}, which is given by
\begin{align}
   Z^{Sp(2)}_\text{extra} = &PE \Bigg[ \frac{- s^2}{(1-s u)(1-s u^{-1})(1-s v)(1-s v^{-1})}\nonumber\\
   &\times\Bigg(\frac{\chi_{\bf 128}^{SO(16)}(\tilde{y}_i)q + \chi_{\bf 120}^{SO(16)}(\tilde{y}_i) q^2}{1-q^2}+     \frac{(s+s^{-1})(u + u^{-1} + v + v^{-1})q^2}{2(1-q^2)}\Bigg)\Bigg].
\end{align}
The instanton partition function is given by
\begin{align}
  Z^{Sp(2)}_\text{inst} = Z^{Sp(2)}_\text{ADHM} / Z^{Sp(2)}_\text{extra}
\end{align}
where $Z^{Sp(2)}_{\rm ADHM}$ is the index computed from the ADHM quantum mechanics 
for $Sp(2)$ instantons. See \cite{Hwang:2014uwa} for the computation of $Z^{Sp(2)}_{\rm ADHM}$, 
which uses the quantum mechanical version of the contour integral formula using Jeffrey-Kirwan 
residues. 

Let us consider the full index $Z^{Sp(2)}\equiv Z^{Sp(2)}_{\rm pert}Z^{Sp(2)}_{\rm inst}$.
$Z^{Sp(2)}$ is very complicated in general.
However, setting $m = \epsilon_+$, $Z^{Sp(2)}$ simplifies a lot and reduces to 
\begin{align}\label{factorization2}
    Z^{Sp(2)}(\tilde{w}_1, \tilde{w}_2,q,s,u,\tilde{y}_i,v=s) = 
    Z^{Sp(1)} (\tilde{w}_1,q,s,u,\tilde{y}_i) Z^{Sp(1)} (\tilde{w}_2,q,s,u,\tilde{y}_i),
\end{align}
where $Z^{Sp(1)}$ is the partition function for the 5d $Sp(1)$ gauge theory obtained by 
compactifying the rank $1$ 6d SCFT for one M5 and M9. ($Z^{Sp(1)}$ is computed by the same 
procedure as explained in the previous paragraph. See \cite{Hwang:2014uwa} for the details.) 
This factorization was checked up to $q^2$ and $\tilde{w}_{1,2}^5$ order. 

\begin{figure}[h!]
 \centering
	\includegraphics[width=\textwidth]{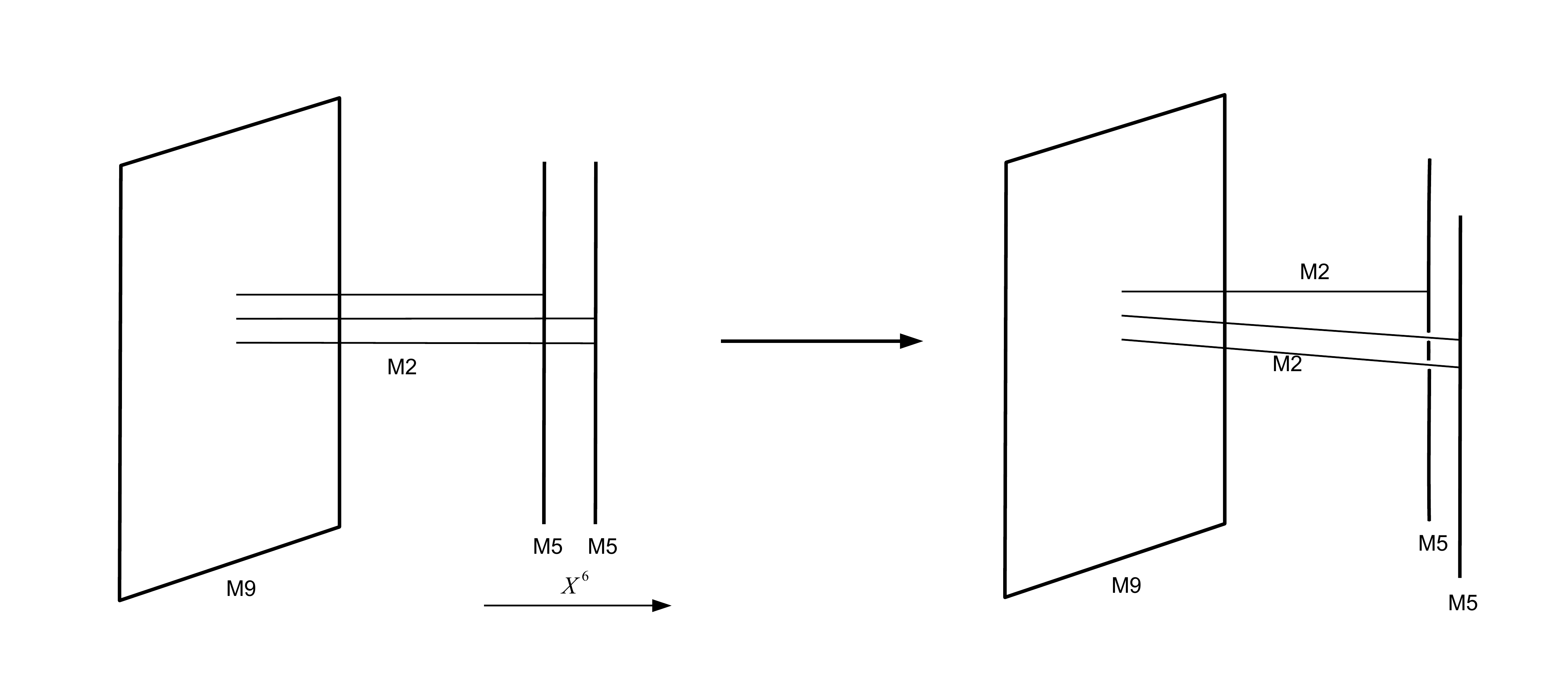}
 \caption{String charge sector $(3,2)$ for the configuration of two M5 branes probing an M9 wall. In the massless case the strings can recombine and one arrives at a configuration of a single E-string from the first M5 brane and two E-strings from the second.}
 \label{fig:M9-M5-massless}
\end{figure}

Now we would like to connect the above findings to the 2d calculus of the previous 
subsection. The two indices are essentially the same, but the latter captures the contributions 
only with positive winding numbers (or 5d electric charges). On the other hand, the former 
also captures some $Sp(2)$ neutral states' contribution with instanton number only. 
The missing part in the 2d calculus can be supplemented by multiplying a $U(1)$ instanton 
partition function factor for 5d maximal SYM, for each M5-brane \cite{Haghighat:2013gba}. 
This factor is given by
\begin{equation}
  Z^{U(1)}=\text{PE}\Bigg[\frac{s(-u-u^{-1}+v+v^{-1})}{(1-su)(1-su^{-1})}\frac{q^2}{1-q^2} \Bigg].
\end{equation}
Therefore, setting $v=s$, the E-string elliptic genera $I_{(n,0)}$ are given by
\begin{equation}
  Z^{U(1)}\sum_{n=0}^\infty w^nI_{(n,0)}(q,s,u,y_i)=Z^{Sp(1)}(\tilde{w},q,s,u,\tilde{y}_i)
\end{equation}
where $I_{(0,0)}\equiv 1$. The coefficients $I_{(n,0)}$ computed from the two different 
approaches (2d gauge theory and instanton calculus) were shown to agree with each other 
\cite{Kim:2014dza}, for $n\leq 4$. The elliptic genera $I_{(n_1,n_2)}$ for the $(n_1,n_2)$ 
strings can be computed from the instanton calculus by
\begin{equation}\label{Sp2}
  \left(Z^{U(1)}\right)^2\sum_{n_1,n_2=0}^\infty w_1^{n_1-n_2}w_2^{n_2}
  I_{(n_1,n_2)}(q,s,u,y_i)=Z^{Sp(2)}(\tilde{w}_1,\tilde{w}_2,q,s,u,\tilde{y}_i)\ .
\end{equation}
One can show that the right hand side of (\ref{Sp2}), computed up to $q^2$ and 
$\tilde{w}_{1,2}^5$ orders from the instanton calculus, completely agrees with 
$I_{(n_1,n_2)}$ computed in the previous subsection for $(n_1,n_2)=(1,1)$, $(1,2)$, $(2,1)$.
In particular, our proposal (\ref{factorization}) is justified from the factorization 
(\ref{factorization2}) of the $Sp(2)$ instanton partition function at $m=\epsilon_+$. We show this result pictorially in Figre \ref{fig:M9-M5-massless}.

\section{D5 branes probing ADE singularities}
\label{sec:D5}
In this section we study a third class of theories that arises from F-theory compactified on an elliptic Calabi-Yau threefold $ X $ defined as follows: we take the base to be the blown-up ALE singularity of ADE type, and over each  blown up $ \mathbb{P}^1 $ we let the elliptic fiber have Kodaira degeneration $ I_N $. Equivalently, we can interpret this setup as Type IIB string theory with N D5 branes probing an ALF singularity of ADE type. We soon will be interested in decompactifying the circle at infinity of ALF and recover the ALE singularity. As follows from the Douglas-Moore construction \cite{Douglas:1996sw}, the resulting $ \mathcal{N}=(1,0) $ six dimensional theory is captured by an affine quiver of ADE type, with the following field content: to a node $ i $ of the affine quiver with Coxeter label $ d_i $ is associated a gauge group $ SU(d_i N) $, and to each edge is associated bifundamental matter. Furthermore, to each node of the quiver corresponds an abelian tensor multiplet. Naively one would expect to have gauge groups $ U(d_i\, N) $, but one finds that in fact the abelian factor $\prod_i U(1)$ is Higgsed via a Green-Schwarz mechanism (apart from a decoupled global U(1) factor) \cite{Blum:1997mm}. We now take the ALF $\to$ ALE limit, so that the the D5 brane associated to the affine node becomes noncompact and gives rise to a global (as opposed to gauge) $ SU(N) $ symmetry. In fact, in the $ A_r $ case, one can actually associate distinct $ SU(N) $ global symmetries to the two non-compact half--$ \mathbb{P}^1 $'s that arise from the affine node.

\begin{figure}[h!]
 \centering
	\includegraphics[width=\textwidth]{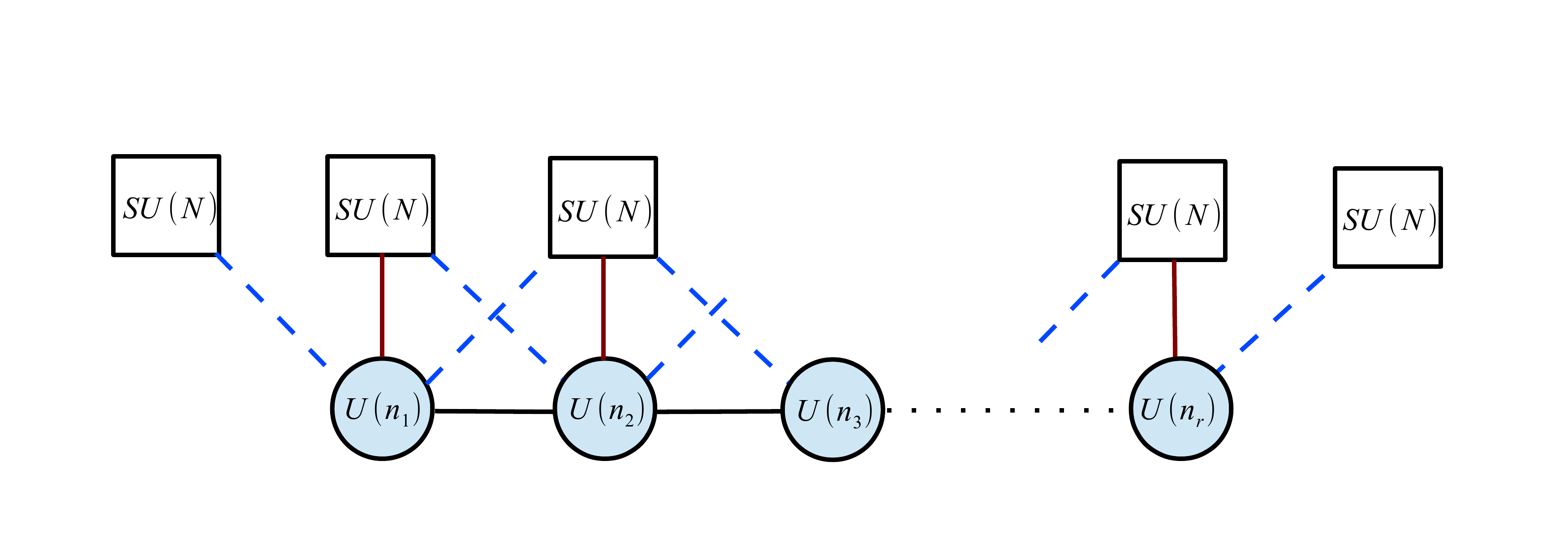}
 \caption{Quiver gauge theory for the non-critical strings of $ N $ D5 branes probing an $A_r$ singularity. Note that the global symmetry are obtained from the affine $ \widehat A_r $ quiver by `opening up' the quiver at the affine node.}
 \label{fig:ArQuiver}
\end{figure}

Of even more interest to us is the quiver gauge theory associated to the self-dual strings of the theory, which arise in F-theory as D3 branes wrapping the blown-up $\mathbb{P}^1$s, or equivalently in Type IIB string theory as D1 branes probing the singularity. The resulting two-dimensional quiver theory was derived for the $ A_N $ singularity by following a suitable generalization of Douglas-Moore construction \cite{Okuyama:2005gq}, and is pictured in Figure \ref{fig:ArQuiver} (this quiver is equivalent to the one depicted in Figure \ref{fig:SuChain}). Note that there is no restriction on the ranks of the gauge groups; the rank at any node is equal to the number of D3 branes (self-dual strings) coupled to the corresponding tensor multiplet.

To avoid clutter, in this section, we work with the exponentiated variables,
\begin{equation}
t= e^{2 \pi i\epsilon_1},\,\, d= e^{2 \pi i \epsilon_2},\,\, c=e^{2 \pi im}.
\end{equation}
We also use exponentiated fugacity variables for gauge and global symmetries, so that, for example, $\theta_1(z) = i q^{1/8}(z^{-1/2}-z^{1/2})\prod_{i=1}^\infty (1-q^i)(1-q^i z)(1-q^i z^{-1}).$\newline

With that, let us first consider an $A_r$ singularity probed by a single D5 brane, that is we set $N=1$.  
The elliptic genus for this case, in the $ (n_1,\dots,n_r) $ sector corresponding to $n_1 $ strings associated to the first node, $ n_2 $ for the second node, and so on, is computed by a contour integral which we write schematically as
\begin{eqnarray}
Z^{A_r,1}_{n_1,\dots,n_r} & = &
\oint\left(\prod_{i=1}^{r}\frac{1}{n_i!}\prod_{a=1}^{n_{i}}\frac{\eta^2 dz_{a}^{(i)}}{2\pi iz_{a}^{(i)}}\right)
\left(\prod_{i=1}^r\prod_{\substack{a,b=1\\a\neq b}}^{n_i}\frac{\theta_1(z_{a}^{(i)}/z_{b}^{(i)})}{\eta}\right)
\left(\prod_{i=1}^r\prod_{a,b=1}^{n_i}\frac{\eta\,\,\theta_1(td\frac{z_{b}^{(i)}}{z_{a}^{(i)}})}{\theta_1(t\frac{z_{a}^{(i)}}{z_{b}^{(i)}})\theta_1(d\frac{z_{a}^{(i)}}{z_{b}^{(i)}})}\right)\nonumber\\
& \times &\left(\prod_{i=1}^{r-1}\prod_{a=1}^{n_i}\prod_{b=1}^{n_{i+1}}\frac{\theta_1(c\sqrt{\frac{d}{t}}\frac{z_{b}^{(i+1)}}{z_{a}^{(i)}})}{\theta_1({c\sqrt{td}\frac{z_{b}^{(i+1)}}{z_{a}^{(i)}}})}\right)
\left(\prod_{i=2}^r\prod_{a=1}^{n_i}\prod_{b=1}^{n_{i-1}}\frac{\theta_1(c\sqrt{\frac{t}{d}}\frac{z_{a}^{(i)}}{z_{b}^{(i-1)}})}{\theta_1(\frac{c}{\sqrt{td}}\frac{z_{a}^{(i)}}{z_{b}^{(i-1)}})}\right)\nonumber\\
& \times &
 \left(\prod_{i=1}^r\prod_{a=1}^{n_{i}}\frac{\theta_1(cz_{a}^{(i)})\theta_1(c\frac{1}{z_{a}^{(i)}})}{\theta_1(\sqrt{td}z_{a}^{(i)})\theta_1(\sqrt{td}\frac{1}{z_{a}^{(i)}})}\right).
\end{eqnarray}
The poles of the integral are labeled by collections of $r$ Young
diagrams $\mathcal Y = \{Y^{(i)}\}_{i=1}^{r}$, such that $ \vert Y^{(i)}\vert = n_i $, and are located at
\begin{equation}
z_{a}^{(i)}=t^{x+\frac{1}{2}}d^{y+\frac{1}{2}}
\end{equation}
where $(x,y)$ are the coordinates of $a$-th box in $Y^{(i)}$ (for example, the two boxes in the diagram $ \square\!\square $ have coordinates (0,0) and (1,0)). Evaluating the residues, we get:

\begin{eqnarray}
&&Z^{A_r,1}_{n_1,\dots,n_r} = \sum_{\mathcal Y}\left(\prod_{i=1}^{r}\prod_{\substack{(x_1,y_1) \in Y^{(i)}\\(x_2,y_2) \in Y^{(i)}}}\frac{\eta\theta_1(t^{x_1-x_2+1}d^{y_1-y_2+1})}{\theta_1(t^{x_1-x_2+1}d^{y_1-y_2})\theta_1(t^{x_1-x_2}d^{y_1-y_2+1})}\right)\nn\\
&&\times\left(\prod_{i=1}^{r}\prod_{\substack{(x_1,y_1) \in Y^{(i)}\\(x_2,y_2) \in Y^{(i)}\\(x_1,y_1)\neq(x_2,y_2)}}\frac{\theta_1(t^{x_1-x_2}d^{y_1-y_2})}{\eta}\right)\nn\\
&  & \times \left(\prod_{i=1}^{r-1}\prod_{\substack{(x_1,y_1) \in Y^{(i+1)}\\(x_2,y_2) \in Y^{(i)}}}\frac{\theta_1(ct^{x_1-x_2-\frac{1}{2}}d^{y_1-y_2+\frac{1}{2}})}{\theta_1(ct^{x_1-x_2+\frac{1}{2}}d^{y_1-y_2+\frac{1}{2}})}\right)\left(\prod_{i=2}^{r}\prod_{\substack{(x_1,y_1) \in Y^{(i)}\\(x_2,y_2) \in Y^{(i-1)}}}\frac{\theta_1(ct^{x_1-x_2+\frac{1}{2}}d^{y_1-y_2-\frac{1}{2}})}{\theta_1(ct^{x_1-x_2-\frac{1}{2}}d^{y_1-y_2-\frac{1}{2}})}\right)\nn\\
&  & \times \left(\prod_{i=1}^{r}\prod_{(x_1,y_1) \in Y^{(i)}}\frac{\theta_1(ct^{x_1+\frac{1}{2}}d^{y_1+\frac{1}{2}})\theta_1(ct^{-x_1-\frac{1}{2}}d^{-y_1-\frac{1}{2}})}{\theta_1(t^{x_1+1}d^{y_1+1})\eta}\right)\left(\prod_{i=1}^{r}\prod_{\substack{(x_1,y_1) \in Y^{(i)}\\(x_1,y_1)\neq (0,0)}}\frac{\eta}{\theta_1(t^{-x_1}d^{-y_1})}\right).\nn\\
\end{eqnarray}
The computation for an arbitrary number $ N $ of D5 branes is only slightly more involved; the flavor symmetry group now also includes a $ SU(N)^{r+2} $ factor; for the factor of $SU(N)$ corresponding to the $ i $-th node, we introduce fugacities $ s_b^{(i)} ,\:b=1,\ldots,N$. These fugacities obey $\prod s_{b}^{(i)}=1$. We label the leftmost and rightmost flavor symmetry nodes in Figure \ref{fig:ArQuiver} respectively by $ i = 0 $ and $ i=r+1 $.\footnote{In the ALE limit, we are free to assign independent fugacities to these two flavor symmetry groups} The elliptic genus is now given by:
\begin{eqnarray}
Z^{A_r,N}_{n_1,\dots,n_r} & = &
\oint\left(\prod_{i=1}^{r}\frac{1}{n_i!}\prod_{a=1}^{n_{i}}\frac{\eta^2 dz_{a}^{(i)}}{2\pi iz_{a}^{(i)}}\right)
\left(\prod_{i=1}^r\prod_{\substack{a,b=1\\a\neq b}}^{n_i}\frac{\theta_1(z_{a}^{(i)}/z_{b}^{(i)})}{\eta}\right)
\left(\prod_{i=1}^r\prod_{a,b=1}^{n_i}\frac{\eta\,\,\theta_1(td\frac{z_{b}^{(i)}}{z_{a}^{(i)}})}{\theta_1(t\frac{z_{a}^{(i)}}{z_{b}^{(i)}})\theta_1(d\frac{z_{a}^{(i)}}{z_{b}^{(i)}})}\right)\nonumber\\
& \times &\left(\prod_{i=1}^{r-1}\prod_{a=1}^{n_i}\prod_{b=1}^{n_{i+1}}\frac{\theta_1(c\sqrt{\frac{d}{t}}\frac{z_{b}^{(i+1)}}{z_{a}^{(i)}})}{\theta_1({c\sqrt{td}\frac{z_{b}^{(i+1)}}{z_{a}^{(i)}}})}\right)
\left(\prod_{i=2}^r\prod_{a=1}^{n_i}\prod_{b=1}^{n_{i-1}}\frac{\theta_1(c\sqrt{\frac{t}{d}}\frac{z_{a}^{(i)}}{z_{b}^{(i-1)}})}{\theta_1(\frac{c}{\sqrt{td}}\frac{z_{a}^{(i)}}{z_{b}^{(i-1)}})}\right)\nonumber\\
 & \times & \left(\prod_{i=1}^r\prod_{a=1}^{n_i}\prod_{b=1}^{N}\frac{\theta_1(c\frac{z_{a}^{(i)}}{s_{b}^{(i-1)}})\theta_1(c \frac{s_{b}^{(i+1)}}{z_{a}^{(i)}})}{\theta_1(\sqrt{td}\frac{z_{a}^{(i)}}{s_{b}^{(i)}})\theta_1(\sqrt{td}\frac{s_{b}^{(i)}}{z_{a}^{(i)}})}\right).
\end{eqnarray}
The poles of this integral are classified by collections $ \mathcal Y $ of colored Young diagrams: $ \mathcal Y = \{\{Y^{(i)}_{\ell_i} \}_{\ell_i=1}^{N}\}_{i=1}^r $, subject to the constraint $\sum_{\ell_i=1}^{N}\vert Y_{\ell_i}^{(i)} \vert=n_{i}$. The pole associated to the $ a$-box in the Young diagram $ Y_{\ell_i}^{(i)} $, with coordinates  $ (x,y) $, is at
\begin{equation}
z_{a}^{(i)}=s_{\ell_i}^{(i)}t^{\frac{1}{2}+x}d^{\frac{1}{2}+y}.
\end{equation}
Evaluating the residues we find:
\begin{eqnarray}
&& Z^{A_r,N}_{n_1,\dots,n_r} =\sum_{\mathcal Y}\left(\prod_{i=1}^r \prod_{\ell,m=1}^{N} \prod_{\substack{(x_1,y_1) \in Y^{(i)}_\ell\\(x_2,y_2) \in Y^{(i)}_m}}\frac{\theta_1(t^{{x_1}-{x_2}}d^{{y_1}-{y_2}})\theta_1(\frac{s_{\ell}^{(i)}}{s_{m}^{(i)}}t^{{x_1}-{x_2}+1}d^{{y_1}-{y_2}+1})}{\theta_1(\frac{s_{\ell}^{(i)}}{s_{m}^{(i)}}t^{{x_1}-{x_2}+1}d^{{y_1}-{y_2}})\theta_1(\frac{s_{\ell}^{(i)}}{s_{m}^{(i)}}t^{{x_1}-{x_2}}d^{{y_1}-{y_2}+1})}\right)
\nonumber\\
& &\times  \left(\prod_{i=1}^{r-1} \prod_{\ell,m=1}^{N}\prod_{\substack{(x_1,y_1)\in Y^{(i)}_\ell\\(x_2,y_2) \in Y^{(i+1)}_m}}\frac{\theta_1(c^{-1}\frac{s_{\ell}^{(i)}}{s_{m}^{(i+1)}}t^{{x_1}-{x_2}+\frac{1}{2}}d^{{y_1}-{y_2}-\frac{1}{2}})}{\theta_1(c^{-1}\frac{s_{\ell}^{(i)}}{s_{m}^{(i+1)}}t^{{x_1}-{x_2}-\frac{1}{2}}d^{{y_1}-{y_2}-\frac{1}{2}})}\right)
\nonumber\\
& &\times  \left(\prod_{i=2}^{r} \prod_{\ell,m=1}^{N}\prod_{\substack{(x_1,y_1)\in Y^{(i)}_\ell\\(x_2,y_2) \in Y^{(i-1)}_m}}\frac{\theta_1(c\frac{s_{\ell}^{(i)}}{s_{m}^{(i-1)}}t^{{x_1}-{x_2}+\frac{1}{2}}d^{{y_1}-{y_2}-\frac{1}{2}})}{\theta_1(c\frac{s_{\ell}^{(i)}}{s_{m}^{(i-1)}}t^{{x_1}-{x_2}-\frac{1}{2}}d^{{y_1}-{y_2}-\frac{1}{2}})}\right)\times
\nonumber
\end{eqnarray}
\begin{eqnarray}
& &\times  \left(\prod_{i=1}^r \prod_{\ell,m=1}^{N} \prod_{(x,y) \in Y^{(i)}_\ell}\frac{\theta_1(c\frac{s_{\ell}^{(i)}}{s_{m}^{(i-1)}}t^{x+\frac{1}{2}}d^{y+\frac{1}{2}})\theta_1(c\frac{s_{m}^{(i+1)}}{s_{\ell}^{(i)}}t^{-x-\frac{1}{2}}d^{-y-\frac{1}{2}})}{\theta_1(\frac{s_{\ell}^{(i)}}{s_{m}^{(i)}}t^{x+1}d^{y+1})\theta_1(\frac{s_{m}^{(i)}}{s_{\ell}^{(i)}}t^{-x}d^{-y}) }\right).
\end{eqnarray}
Note that the products include various factors of $ \theta_1(1) $, which however completely cancel against each other.\newline

\begin{figure}[h!]
 \centering
	\includegraphics[width=\textwidth]{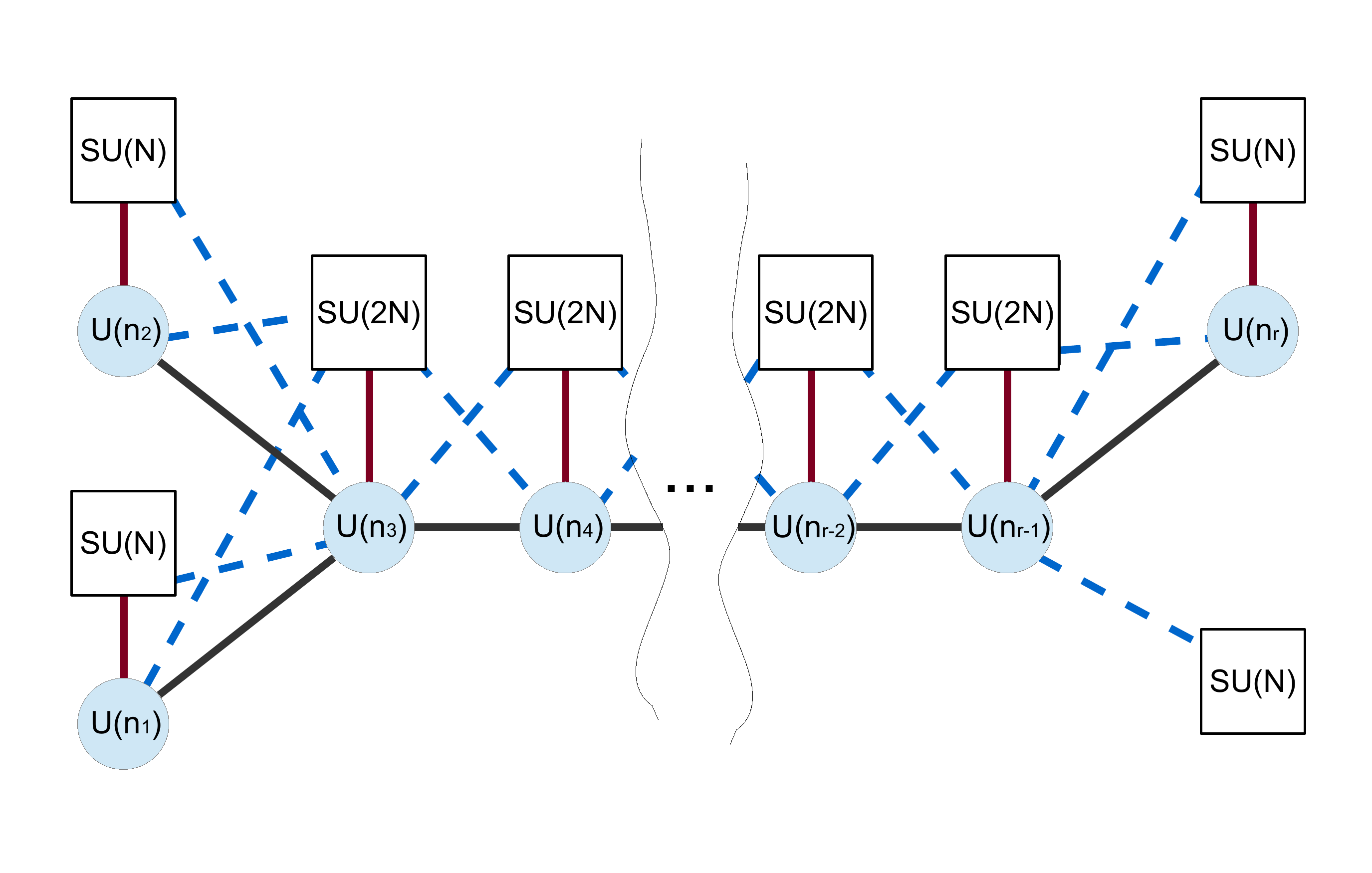}
 \caption{Quiver gauge theory for the non-critical strings of $ N $ D5 branes probing a $D_r$ singularity.}
 \label{fig:DrQuiver}
\end{figure}
\begin{figure}[h!]
  \centering
	\includegraphics[width=0.4\textwidth]{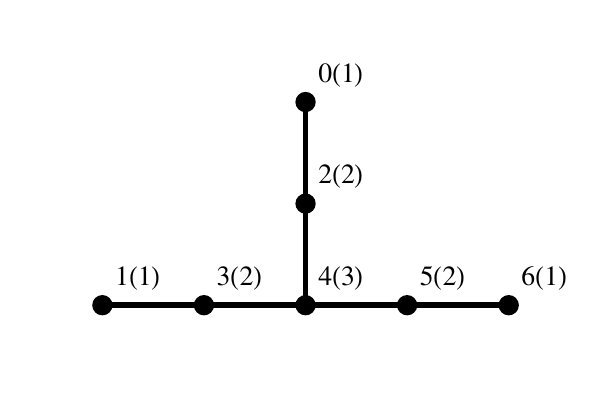}
  \caption{Oriented $\widehat{E}_6$ quiver with a particular labelling of the nodes $i$. The numbers in the parentheses represent the Coxeter labels $d_i$.}
  \label{fig:AffineE6quiver}
\end{figure}
\begin{figure}[h!]
 \centering
	\includegraphics[width=\textwidth]{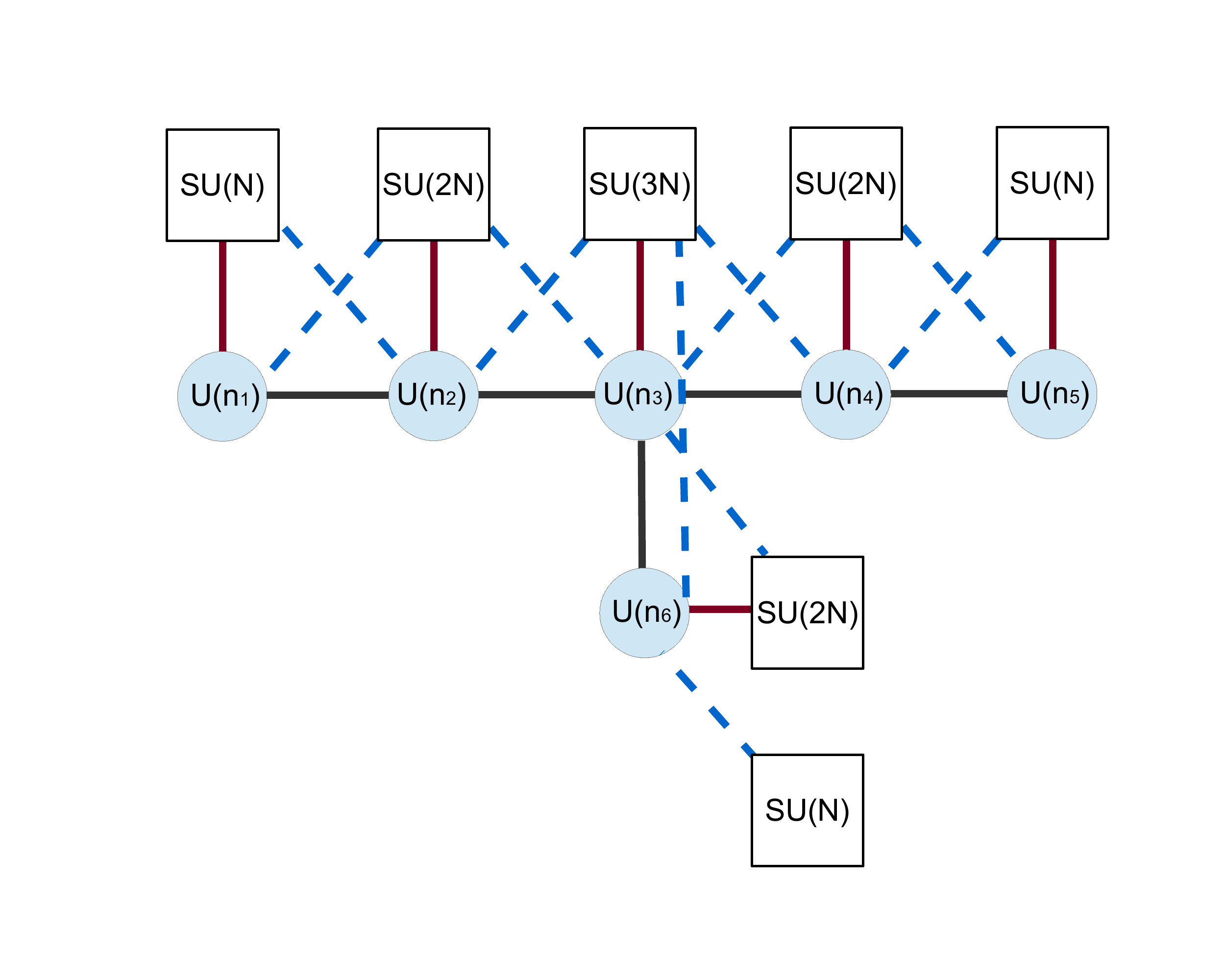}
\caption{Quiver gauge theory for the non-critical strings of $ N $ D5 branes probing an $E_6$ singularity.}

 \label{fig:E6Quiver}
\end{figure}
\begin{figure}[h!]
 \centering
	\includegraphics[width=0.8\textwidth]{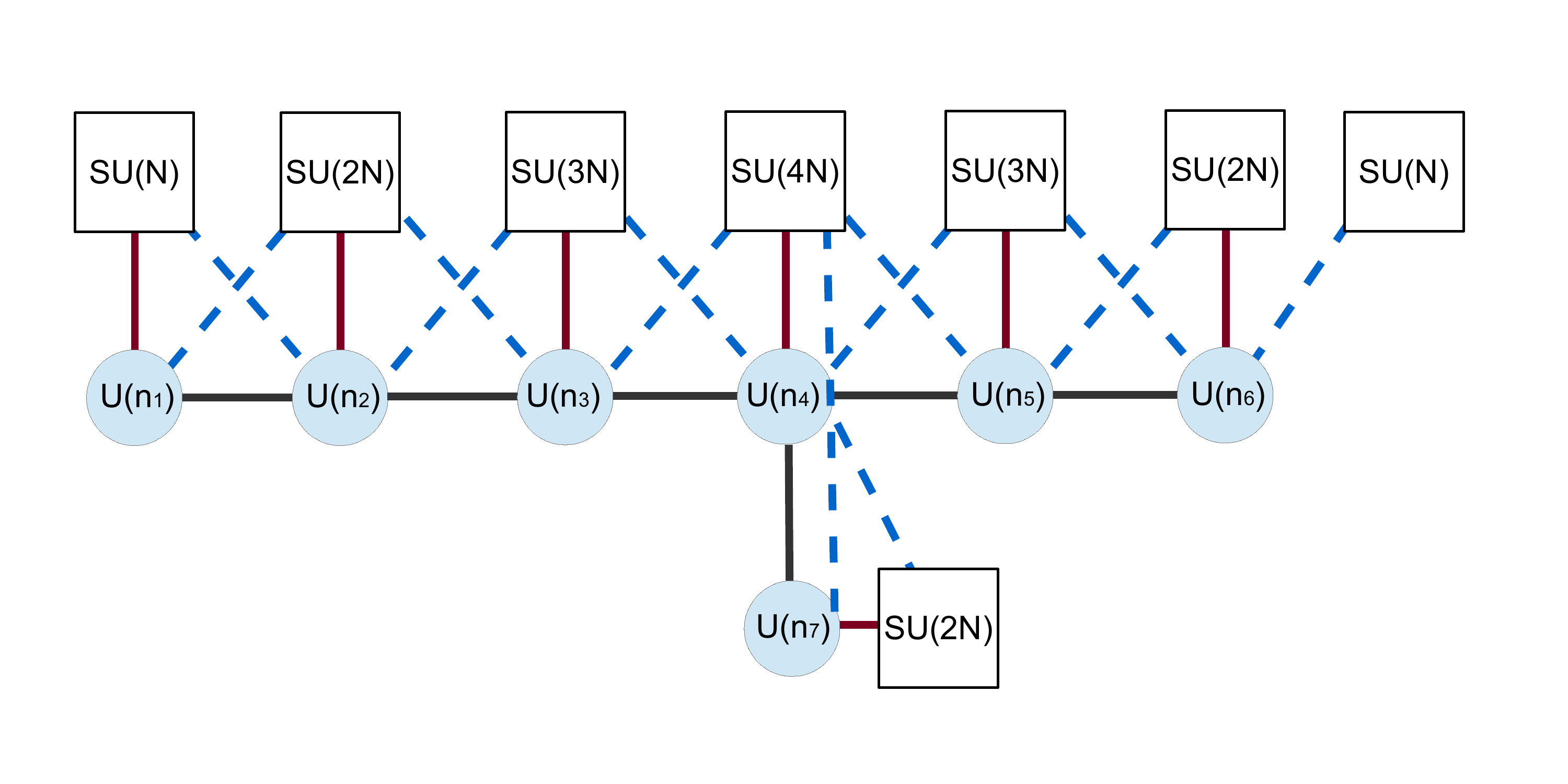}
 \caption{Quiver gauge theory for the non-critical strings of $ N $ D5 branes probing an $E_7$ singularity.}
 \label{fig:E7Quiver}
\end{figure}
\begin{figure}[h!]
 \centering
	\includegraphics[width=\textwidth]{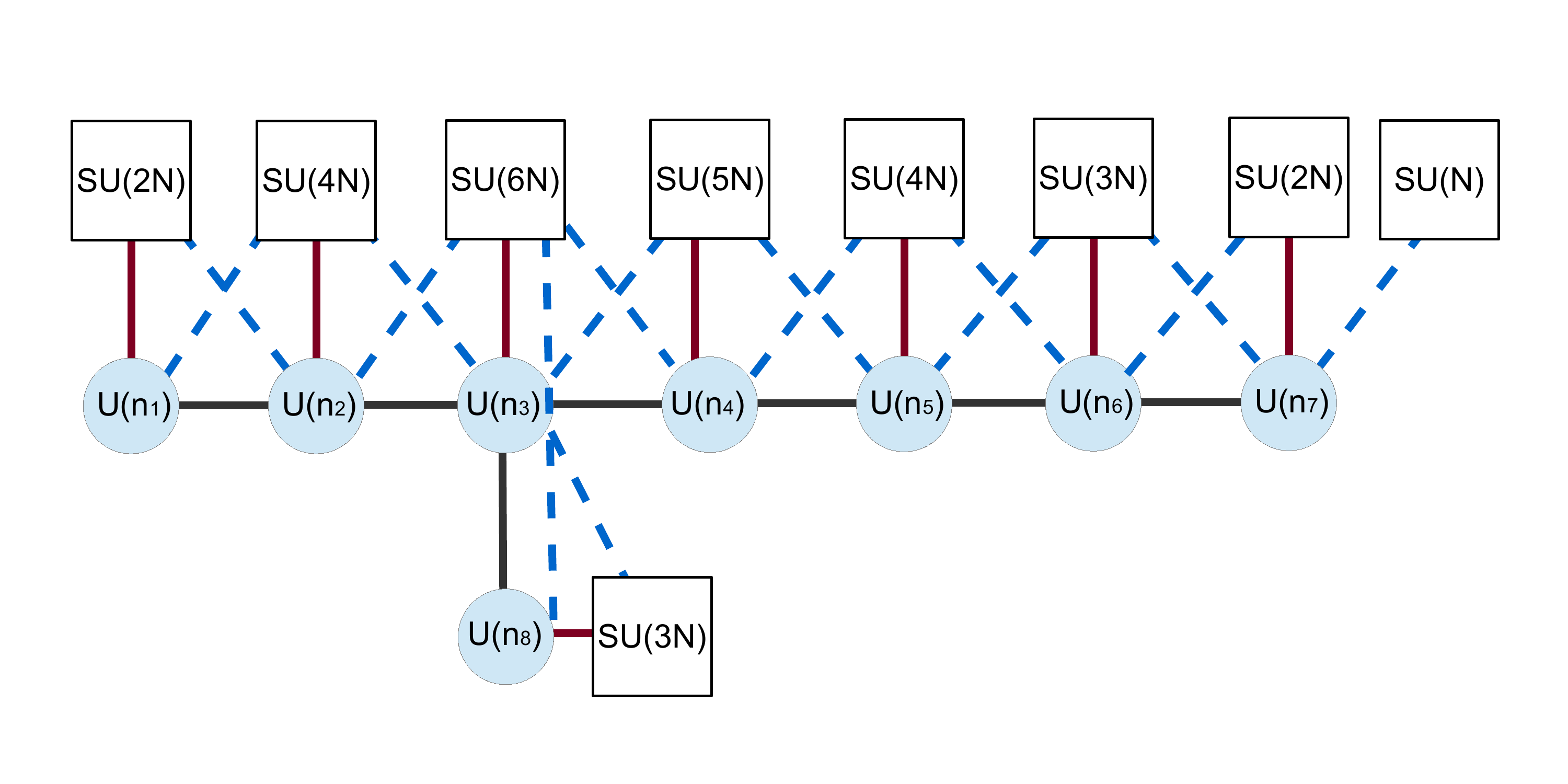}
 \caption{Quiver gauge theory for the non-critical strings of $ N $ D5 branes probing an $E_8$ singularity.}
 \label{fig:E8Quiver}
\end{figure}

Finally we pass to the case of $N$ D5 branes probing singularities of type $D_r$ or $E_6,E_7,E_8$. Their two-dimensional quivers are pictured in Figures \ref{fig:DrQuiver}\,--\ref{fig:E8Quiver}. An important difference between the $A$ case and the $D$ and $E$ cases is that in the latter the $U(1)_m$ symmetry is anomalous. Hence, we will not be able to refine the index with the corresponding fugacity $c$.
The Lagrangian of the $(0,4)$ theory is chiral. This is apparent from the fact that $\Lambda^Q$ and $\Lambda^{\tilde Q}$ transform oppositely. The same is the case with the fields $\Lambda^B,\Lambda^{\tilde B}, \Sigma$ and $\Phi$. Although, for $D$ and $E$ type quivers there is no  preferred orientation of, say, the $\Phi$ arrow. But as it will soon become clear, the elliptic genus of the theory does not depend on the choice of this orientation.

Let us label the gauge nodes by an index $i$ taking the values from $1$ to $r$. The range of the index for the flavor nodes, on the other hand, runs from $ 0 $ to $ r $, where $ i=0 $ corresponds to the affine node. Let $s_{b}^{(i)},\:b=1,\ldots,d_{i}N$ be the fugacity corresponding to the flavor symmetry group $SU(d_{i}N)$ where $d_{i}$ is the Coxeter label of the node $i$. These fugacities obey $\prod s_{b}^{(i)}=1$. Let $M_{ij}$ be the adjacency matrix of the quiver with \emph{some} orientation. 

For example, Figure \ref{fig:AffineE6quiver} shows the Coxeter labels of the affine $E_6$ quiver together with a particular labeling. The corresponding adjacency matrix is given by
\begin{equation}
	M^{\widehat{E}_6} = \left(\begin{array}{ccccccc}
		0 & 0 & 1 & 0 & 0 & 0 & 0\\
		0 & 0 & 0 & 1 & 0 & 0 & 0\\
		0 & 0 & 0 & 0 & 1 & 0 & 0\\
		0 & 0 & 0 & 0 & 1 & 0 & 0\\
		0 & 0 & 0 & 0 & 0 & 1 & 0\\
		0 & 0 & 0 & 0 & 0 & 0 & 1\\
		0 & 0 & 0 & 0 & 0 & 0 & 0
	\end{array}\right).
\end{equation}
The elliptic genus of the $2d$ theory for $ G = D_r,E_6,E_7,E_8 $ is then given by
\begin{eqnarray}
Z^{G,N}_{n_1,\dots,n_r} & = &
\oint\left(\prod_{i=1}^{r}\frac{1}{n_i!}\prod_{a=1}^{n_{i}}\frac{\eta^2 dz_{a}^{(i)}}{2\pi iz_{a}^{(i)}}\right)
\left(\prod_{i=1}^r\prod_{\substack{a,b=1\\a\neq b}}^{n_i}\frac{\theta_1(z_{a}^{(i)}/z_{b}^{(i)})}{\eta}\right)
\left(\prod_{i=1}^r\prod_{a,b=1}^{n_i}\frac{\eta\,\,\theta_1(td\frac{z_{b}^{(i)}}{z_{a}^{(i)}})}{\theta_1(t\frac{z_{a}^{(i)}}{z_{b}^{(i)}})\theta_1(d\frac{z_{a}^{(i)}}{z_{b}^{(i)}})}\right)\nonumber\\
& \times & \left(\prod_{i=1}^r\prod_{a=1}^{n_i}\prod_{b=1}^{N d_i}\frac{\eta^2}{\theta_1(\sqrt{td}\frac{z_{a}^{(i)}}{s_{b}^{(i)}})\theta_1(\sqrt{td}\frac{s_{b}^{(i)}}{z_{a}^{(i)}})}\right)\left(\prod_{j=0}^r\prod_{b=1}^{d_j N}\left(\frac{\theta_1(\frac{z_{a}^{(i)}}{s_{b}^{(j)}})}{\eta}\right)^{M_{ij}}\left(\frac{\theta_1(\frac{s_{b}^{(j)}}{z_{a}^{(i)}})}{\eta}\right)^{M_{ji}}\right)\nonumber
\\
& \times & \left(\prod_{i,j=1}^r \prod_{a=1}^{n_i} \prod_{b=1}^{n_j} \left(\frac{\theta_1(\sqrt{\frac{d}{t}}\frac{z_{a}^{(i)}}{z_{b}^{(j)}})\theta_1( \sqrt{\frac{t}{d}}\frac{z_{a}^{(i)}}{z_{b}^{(j)}})}{\theta_1(\frac{1}{\sqrt{td}}\frac{z_{b}^{(j)}}{z_{a}^{(i)}})\theta_1(\frac{1}{\sqrt{td}}\frac{z_{a}^{(i)}}{z_{b}^{(j)}})}\right)^{M_{ij}}\right). \nonumber \\
\label{eq:I2da} 
\end{eqnarray}
Using the identity $\theta_1(x)=-\theta_1(x^{-1})$ we get (up to a sign):
\begin{eqnarray}
Z^{G,N}_{n_1,\dots,n_r} & = &\!
\oint\left(\prod_{i=1}^{r}\frac{1}{n_i!}\prod_{a=1}^{n_{i}}\frac{\eta^2 dz_{a}^{(i)}}{2\pi iz_{a}^{(i)}}\right)\!
\left(\prod_{i=1}^r\prod_{\substack{a,b=1\\a\neq b}}^{n_i}\frac{\theta_1(z_{a}^{(i)}/z_{b}^{(i)})}{\eta}\right)\!
\left(\prod_{i=1}^r\prod_{a,b=1}^{n_i}\frac{\eta\,\,\theta_1(td\frac{z_{b}^{(i)}}{z_{a}^{(i)}})}{\theta_1(t\frac{z_{a}^{(i)}}{z_{b}^{(i)}})\theta_1(d\frac{z_{a}^{(i)}}{z_{b}^{(i)}})}\right)\nonumber\\
& \times & \left(\prod_{i=1}^r\prod_{a=1}^{n_i}\prod_{b=1}^{N d_i}\frac{\eta^2}{\theta_1(\sqrt{td}\frac{z_{a}^{(i)}}{s_{b}^{(i)}})\theta_1(\sqrt{td}\frac{s_{b}^{(i)}}{z_{a}^{(i)}})}\right)\left(\prod_{i=1}^r\prod_{j=0}^r\prod_{b=1}^{d_j N}\left(\frac{\theta_1(\frac{s_{b}^{(j)}}{z_{a}^{(i)}})}{\eta}\right)^{M_{ij}+M^T_{ij}}\right)\nonumber
\\
& \times & \left(\prod_{i,j=1}^r \prod_{a=1}^{n_i} \prod_{b=1}^{n_j} \left(\frac{\theta_1(\sqrt{\frac{d}{t}}\frac{z_{a}^{(i)}}{z_{b}^{(j)}})}{\theta_1(\frac{1}{\sqrt{td}}\frac{z_{a}^{(i)}}{z_{b}^{(j)}})}\right)^{M_{ij}+M^T_{ij}}\right).\label{eq:I2db}
\end{eqnarray}
As it is clear from the expression, the elliptic genus depends only on the combination  $M_{ij}+M^T_{ij}\equiv A_{ij}$, the \emph{undirected} adjacency matrix of the affine $D$ or $E$ quiver.
The poles of this integral are classified by collections $ \mathcal Y $ of colored Young diagrams: $ \mathcal Y = \{\{Y^{(i)}_{\ell_i} \}_{\ell_i=1}^{d_i N}\}_{i=1}^r $, subject to the constraint $\sum_{\ell_i=1}^{d_{i} N}\vert Y_{\ell_i}^{(i)} \vert=n_{i}$. The pole associated to the $ a$-box in the Young diagram $ Y_{\ell_i}^{(i)} $, with coordinates  $ (x^{(i)}_{\ell_i},y^{(i)}_{\ell_i}) $, is at
\begin{equation}
z_{a}^{(i)}=s_{\ell_i}^{(i)}t^{\frac{1}{2}+x^{(i)}_{\ell_i}}d^{\frac{1}{2}+y^{(i)}_{\ell_i}}.
\end{equation}
Evaluating the residues we find:
\begin{eqnarray}
&&Z^{G,N}_{n_1,\dots,n_r} =  \sum_{\mathcal Y}\left(\prod_{i=1}^r \prod_{\ell,m=1}^{d_i N} \prod_{\substack{(x_1,y_1) \in Y^{(i)}_\ell\\(x_2,y_2) \in Y^{(i)}_m}}\frac{\theta_1(\frac{s_{\ell}^{(i)}}{s_{m}^{(i)}}t^{{x_1}-{x_2}}d^{{y_1}-{y_2}})\theta_1(\frac{s_{\ell}^{(i)}}{s_{m}^{(i)}}t^{{x_1}-{x_2}+1}d^{{y_1}-{y_2}+1})}{\theta_1(\frac{s_{\ell}^{(i)}}{s_{m}^{(i)}}t^{{x_1}-{x_2}+1}d^{{y_1}-{y_2}})\theta_1(\frac{s_{\ell}^{(i)}}{s_{m}^{(i)}}t^{{x_1}-{x_2}}d^{{y_1}-{y_2}+1})}\right)
\nonumber\\
&&\times \left(\prod_{i,j=1}^r \prod_{\ell=1}^{d_i N}\prod_{m=1}^{d_j N}\prod_{\substack{(x_1,y_1)\in Y^{(i)}_\ell\\(x_2,y_2) \in Y^{(j)}_m}}\left(\frac{\theta_1(\frac{s_{\ell}^{(i)}}{s_{m}^{(j)}}t^{{x_1}-{x_2}-\frac{1}{2}}d^{{y_1}-{y_2}+\frac{1}{2}})}{\theta_1(\frac{s_{\ell}^{(i)}}{s_{m}^{(j)}}t^{{x_1}-{x_2}-\frac{1}{2}}d^{{y_1}-{y_2}-\frac{1}{2}})}\right)^{A_{ij}}\right)
\nonumber\\
&& \times\left(\prod_{i=1}^r \prod_{\ell=1}^{d_i N} \prod_{(x,y) \in Y^{(i)}_\ell}\frac{\prod_{j=0}^{r}\prod_{m=1}^{d_j N}\left(\theta_1(\frac{s_{\ell}^{(i)}}{s_{m}^{(j)}}t^{x+\frac{1}{2}}d^{y+\frac{1}{2}})/\eta\right)^{A_{ij}}}{\prod_{m=1}^{d_i N}\theta_1(\frac{s_{\ell}^{(i)}}{s_{m}^{(i)}}t^{x+1}d^{y+1})\theta_1(\frac{s_{m}^{(i)}}{s_{\ell}^{(i)}}t^{-x}d^{-y})/\eta^2}\right)\!.
\end{eqnarray}

\section*{Acknowledgements}
We are grateful to Clay Cordova, Michele Del Zotto, Thomas Dumitrescu, Jonathan Heckman, David Morrison, Tom Rudelius and Shu-Heng Shao for discussions. The work of A.G. is supported by the Raymond and Beverly Sackler Foundation Fund and the NSF grant PHY-1314311. The work of B.H. is supported by the NSF grant DMS-1159412. The work of C.V. is supported in part by NSF grant PHY-1067976. The work of J.K. and S.K. is supported in part by the National Research Foundation of Korea Grants 2012R1A1A2042474, 2012R1A2A2A02046739.
 
\appendix

\section{Quiver for $ N $ small $ E_8 $ instantons with $ m \neq 0 $}
\begin{figure}[htbp]
  \centering
   \includegraphics[width=10cm]{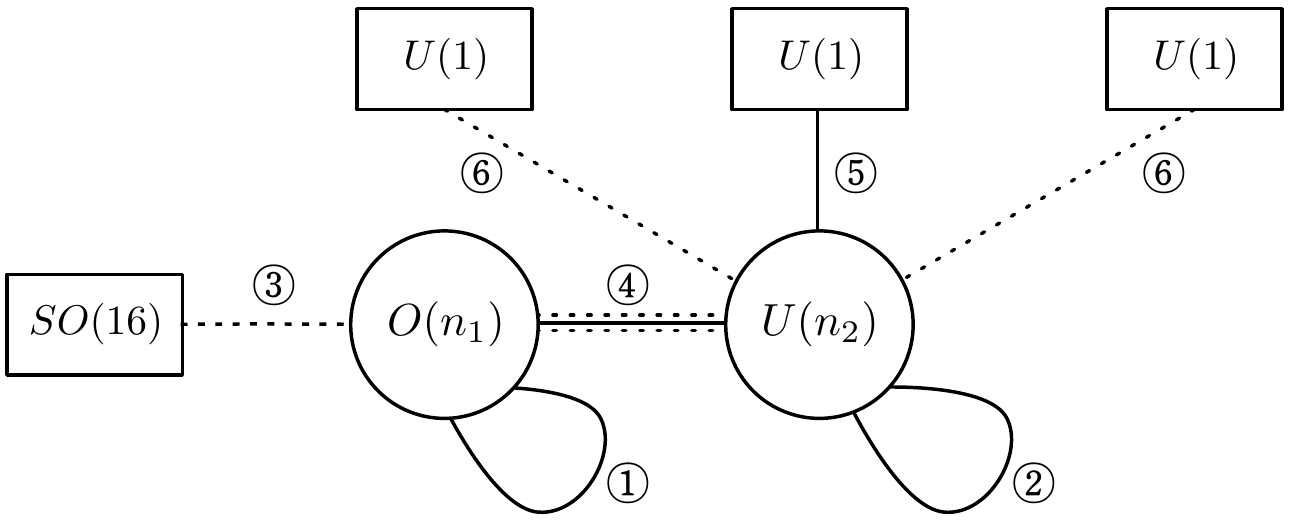}
  \caption{(0,2) quiver for two small $ E_8 $ instantons.}
  \label{fig:quiver}
\end{figure}

In this appendix we present a quiver (Figure \ref{fig:quiver}) which we conjecture to reproduce elliptic genera for the strings of two small $E_8$ instantons with non-zero value of the antisymmetric hypermultiplet mass. The generalization to $ N $ small $ E_8 $ instantons is obtained straightforwardly by adjoining additional nodes with unitary gauge group, as for the M-strings. The results for the elliptic genera of the lowest charge sectors agree with the instanton computation for 5d $Sp(2)$ gauge theory with $8$ fundamental hypermultiplets with masses $m_l$ and one anti-symmetric hypermultiplet of mass $m$. Although the quiver in Figure \ref{fig:quiver} has only manifest $(0,2)$ supersymmetry we conjecture that it flows to a $(0,4)$ theory in the IR.

The matter content of the 2d theory is a combination of that for for E-string and that for M-strings, with $ (0,4) $ bifundamental twisted hypermultiplets and Fermi multiplets; it is summarized in Table \ref{tab:field}. Let us denote the sector corresponding to $ n_1 $ M9-M5 strings and $ n_2 $ M5-M5 strings by $ (n_1, n_2) $. If $ n_1 = 0 $ or $ n_2 = 0 $ the 2d CFT reduces respectively to the one for M-strings or E-strings.  In the following we compute the elliptic genera for a number of other sectors.

\begin{table}[htbp]
\centering
\begin{tabular}{|c||c|c|c|}
  \hline
  Type & Field & Multiplet & Representation \\ \hline 
  1 & $(a^{\alpha}{}_{\dot{\beta}}, \lambda_-^{\alpha A})_{1}$ & hyper. & symm.\\
  2 & $(a^{\alpha}{}_{\dot{\beta}}, \lambda_-^{\alpha A})_{2}$ & hyper. & adjoint\\
  3 & $(\Psi_+)$ & Fermi & fund.\\
  4 & $(\varphi_{A}, \chi_{-}^{\dot{\alpha}})_1 \oplus (\varphi_{A}, \chi_{-}^{\dot{\alpha}})_2$ & twisted hyper. & bifund.\\
  4 & $(\chi_+^{\alpha})_1 \oplus (\chi_+^{\alpha})_2$ & Fermi & bifund. \\
  5 & $(q_{\dot{\alpha}},\psi^{A}_+)$ & hyper. & fund. \\
  6 & $(\psi_-)_1 \oplus (\psi_-)_2$ & Fermi. & fund. \\\hline
\end{tabular}
\caption{Field content of the 2d quiver theory for two small $E_8$ instantons.}
\label{tab:field}
\end{table}

\subsection{The $(n_1, n_2) = (1,1)$ sector}
The gauge group in this case is $ O(1)\times U(1) $. Combining the one-loop determinants, the zero-mode integral $\mathcal{I}_{(1,1)}$ is given by
\begin{align}
  \oint \frac{du}{2} \sum_{i=1}^{4} &\Bigg(\frac{-\eta^2}{\theta_1( \epsilon_1)\theta_1( \epsilon_2)}\Bigg)_1 \Bigg(\frac{\eta^3 \theta_1 (\epsilon_1 + \epsilon_2)}{\theta_1( \epsilon_1)\theta_1( \epsilon_2)}\Bigg)_2  \Bigg(\prod_{l=1}^{8}\frac{\theta_i (m_l)}{\eta}\Bigg)_3  \nonumber\\
 &\Bigg( \frac{\theta_1 ( \pm m + \epsilon_- \pm a_i \mp u)}{\theta_1 (\pm m - \epsilon_+ \pm a_i \mp u)}\Bigg)_{4} \Bigg( \frac{\theta_1 ( m \pm u)}{\theta_1 (\epsilon_+ \pm u )}\Bigg)_{5,6},
\end{align}
where $a_i = (0, \frac{1}{2}, \frac{1+\tau}{2}, \frac{\tau}{2})_i$. Note that we are using compact $\pm$ notation; for example:
\begin{align}
  \theta_1 (\pm m + \epsilon_- \pm a_i \mp u) &\equiv \theta_1 ( m + \epsilon_- + a_i - u) \theta_1 (- m + \epsilon_- - a_i + u)\\
  \theta_1 (m \pm u_1) &\equiv \theta_1 ( m + u_1) \theta_1 (m- u_1),
\end{align}
and so on. The Jeffrey-Kirwan residue operation picks the poles $u = -\epsilon_+$ and $u = m + \epsilon_+ + a_i$. Collecting the residues one finds:
\begin{align}
 \mathcal{I}_{(1,1)} &= \Bigg[\sum_{i=1}^{4} \Bigg(\prod_{l=1}^{8}\frac{\theta_i (m_l)}{\eta}\Bigg)
  \cdot \Bigg(\frac{\theta_1(m + \epsilon_+)\theta_1(m - \epsilon_+)}{\eta^3\theta_1(\epsilon_1 + \epsilon_2) } \frac{\theta_i(m + \epsilon_1)\theta_i(m + \epsilon_2)}{\theta_i(m)\theta_i(m + \epsilon_1 + \epsilon_2)} \\
  &\hspace{.8in}+ \frac{\theta_1 (\epsilon_1)\theta_1 (\epsilon_2)}{\eta^3 \theta_1 (\epsilon_1 + \epsilon_2)}\frac{\theta_i (2m + \epsilon_+)\theta_i (\epsilon_+)}{\theta_i (m + \epsilon_1 + \epsilon_2)\theta_i (m)}\Bigg)\Bigg] \times \frac{-\eta^5 \theta_1 (\epsilon_1 + \epsilon_2)}{2\theta_1(\epsilon_1)^2 \theta_1(\epsilon_2)^2} . \nonumber
\end{align}
We have checked up to  powers of $q^{4}$ (here, $ q = e^{2\pi i \tau} $) that this matches with results from $ Sp(2) $ instanton calculus
\begin{align}
  \mathcal{I}_{(1,1)}= \frac{\eta^2 \theta_1 (m - \epsilon_-) \theta_1 (-m - \epsilon_-)}{2\theta_1(\epsilon_1)^2 \theta_1(\epsilon_2)^2}\sum_{i=1}^{4} \Bigg(\prod_{l=1}^{8}\frac{\theta_i (m_l)}{\eta}\Bigg).
\end{align}

\subsection{The $(n_1, n_2) = (1,2)$ sector}
The zero-mode integral is given by
\begin{align}
  \mathcal{I}_{(1,2)} = & \hspace{.5in}  \oint \frac{du_1 du_2}{4} \; \sum_{i=1}^{4}  \Bigg(\prod_{l=1}^{8}\frac{\theta_i(m_l)}{ \eta}\Bigg)\frac{\theta_1 ( \pm(u_1-u_2)) \theta_1 ( \epsilon_1 + \epsilon_2 \pm(u_1-u_2))}{\theta_1 ( \epsilon_1 \pm(u_1-u_2)) \theta_1 ( \epsilon_2 \pm(u_1-u_2))} \nonumber \\
  &\times \frac{\theta_1(\pm m + \epsilon_- \pm a_i \mp u_{1})}{\theta_1(\pm m - \epsilon_+ \pm a_i \mp u_{1})} \frac{\theta_1(\pm m + \epsilon_- \pm a_i \mp u_{2})}{\theta_1(\pm m - \epsilon_+ \pm a_i \mp u_{2})}  \frac{\theta_1(m \pm u_1)}{\theta_1(\epsilon_+ \pm u_1)}\frac{\theta_1(m \pm u_2)}{\theta_1(\epsilon_+ \pm u_2)}\nonumber\\
  &\times \frac{-\eta^2}{\theta_1( \epsilon_1)\theta_1( \epsilon_2)} \Bigg(\frac{\eta^3 \theta_1 (\epsilon_1 + \epsilon_2)}{\theta_1( \epsilon_1)\theta_1( \epsilon_2)}\Bigg)^{2}, 
\end{align}
where $a_i = (0, \frac{1}{2}, \frac{1+\tau}{2}, \frac{\tau}{2})_i$. The Jeffrey-Kirwan prescription picks the poles at $(u_1, u_2) = (-\epsilon_+, m + \epsilon_+ + a_i)$, $(m + \epsilon_+ + a_i, -\epsilon_+)$, $(-\epsilon_+, -\epsilon_+ - \epsilon_{1,2})$, $(-\epsilon_+ - \epsilon_{1,2}, -\epsilon_+ )$, as well as $(m+\epsilon_+ + a_i, m + \epsilon_+ + a_i -\epsilon_{1,2})$, $(m+\epsilon_+ + a_i -\epsilon_{1,2}, m + \epsilon_+ + a_i )$, where the last two give us vanishing residues. The final result is
\begin{align}
  \mathcal{I}_{(1,2)}&=-\sum_{i=1}^{4}\Bigg(\prod_{l=1}^{8}\frac{\theta_i(m_l)}{ \eta}\Bigg)\frac{\eta^2 \theta_1(m \pm \epsilon_+)}{\theta_1(\epsilon_1)^2 \theta_1(\epsilon_2)^2}\Bigg[\frac{\theta_i(\epsilon_+) \theta_i (4 \epsilon_+ + m) \theta_i (2m + \epsilon_+)}{\theta_i(m) \theta_i (2 \epsilon_1 + \epsilon_2 + m) \theta_i (\epsilon_1 + 2 \epsilon_2 + m)} \nonumber\\
  &\hspace{2.3in}+ \frac{\theta_1(m \pm (\frac{3\epsilon_1}{2} + \frac{\epsilon_2}{2}))}{\theta_1(2 \epsilon_1) \theta_1 (\epsilon_2 - \epsilon_1)} \frac{\theta_i(\epsilon_2 + m)\theta_i (2 \epsilon_1 + m)}{\theta_i (2 \epsilon_1 + \epsilon_2 + m) \theta_i(m)}\nonumber\\
  &\hspace{2.3in}+ \frac{\theta_1(m \pm (\frac{\epsilon_1}{2} + \frac{3\epsilon_2}{2}))}{\theta_1(2 \epsilon_2) \theta_1 (\epsilon_1 - \epsilon_2)} \frac{\theta_i(\epsilon_1 + m)\theta_i (2 \epsilon_2 + m)}{\theta_i (\epsilon_1 + 2\epsilon_2 + m) \theta_i(m)}\Bigg],
\end{align}
which again agrees with results from $ Sp(2) $ instanton calculus.
\subsection{The $(n_1, n_2) = (2,1)$ sector}
For the continuous $O(2)$ holonomy, we have to deal with a rank two contour integration. One finds:
\begin{align}
  \mathcal{I}_{(2,1)}^{0} =&  \oint \frac{du_1 du_2}{2} \; \frac{\eta^7 \theta_1 (\epsilon_1 + \epsilon_2)}{\theta_1( \epsilon_{1,2}) \theta_1 (\epsilon_{1,2} \pm 2 u_1)} \frac{\eta^3 \theta_1 (\epsilon_1 + \epsilon_2)}{\theta_1( \epsilon_1)\theta_1( \epsilon_2)}  \Bigg(\prod_{l=1}^{8}\frac{\theta_1(m_l \pm u_1)}{ \eta}\Bigg) \\
  &\times \frac{\theta_1(\pm m + \epsilon_- + u_1 \mp u_2)}{\theta_1(\pm m - \epsilon_+ + u_1 \mp u_2)} \frac{\theta_1(\pm m + \epsilon_- - u_1 \mp u_2)}{\theta_1(\pm m - \epsilon_+ - u_1 \mp u_2)}  \frac{\theta_1(m \pm u_2)}{\theta_1(\epsilon_+ \pm u_2)}. \nonumber
\end{align}
If we choose $\eta = e_1 + \epsilon \, e_2$ with $\epsilon \ll 1$, the Jeffrey-Kirwan residue prescription picks the  following poles:
\begin{itemize}
  \item $u_1 + u_2 - \epsilon_+ - m =0$, $\epsilon_+ - u_2 = 0$: $(u_1, u_2) = (m , \epsilon_+)$.
  \item $u_2 + \epsilon_+ = 0$, $u_1 - u_2 - \epsilon_+ + m = 0$: $(u_1 , u_2) = (-m, -\epsilon_+)$.
  \item $u_1 + u_2 - \epsilon_+ - m =0$, $u_1 - u_2 - \epsilon_+ + m = 0$: $(u_1, u_2) = (\epsilon_+ + a_i, m + a_i)$.
  \item $2 u_1 + \epsilon_{1,2} = 0$, $u_1 + u_2 - \epsilon_+ - m = 0$: $(u_1, u_2) = \!(-\tfrac{\epsilon_{1,2}}{2}+a_i, \epsilon_{1,2}\!+\tfrac{\epsilon_{2,1}}{2}\!+m-a_i)$.
  \item $2 u_1 + \epsilon_{1,2} = 0$, $u_2 + \epsilon_+ = 0$: $(u_1, u_2) = (-\tfrac{\epsilon_{1,2}}{2}+a_i, -\epsilon_+)$.
  \item $2 u_1 + \epsilon_{1,2} = 0$, $-u_1 + u_2 -\epsilon_+ - m = 0$: $(u_1 , u_2) = (-\tfrac{\epsilon_{1,2}}{2}+a_i, \tfrac{\epsilon_{2,1}}{2}+m+a_i)$.
  \\These poles will give zero residues, due to numerators $\theta_1 (-m + \epsilon_- + u_1 + u_2) \theta_1 (m + \epsilon_- - u_1 - u_2)$.
\end{itemize}
Here, $a_i$ runs over $\{0, \frac{1}{2}, \frac{1+\tau}{2},\frac{\tau}{2}\}$. 
Doing the integration with sign factors dictated by Jeffrey-Kirwan residue prescription, one finds
\begin{small}\begin{align}
  \mathcal{I}_{(2,1)}^{0}=& +\frac{\eta^{-12} \theta_1 (\epsilon_+ \pm m) \prod_{l=1}^{8}\theta_1(m_l \pm m)}{2\theta_1(\epsilon_{1,2}) \theta_1(\epsilon_{1,2} + 2m) \theta_1 (2m) \theta_1(\epsilon_1 + \epsilon_2 - 2m)}\nonumber \\
  & - \frac{\eta^{-12} \theta_1 (\epsilon_+ \pm m) \prod_{l=1}^{8}\theta_1(m_l \pm m)}{2\theta_1(\epsilon_{1,2}) \theta_1(\epsilon_{1,2} - 2m) \theta_1 (2m) \theta_1(\epsilon_1 + \epsilon_2 + 2m)} \nonumber \\
  &+ \sum_{i=1}^{4}\frac{\eta^{-12} \theta_i (0) \theta_i (2m) \prod_{l=1}^{8} \theta_i (m_l \pm \epsilon_+)}{4\theta_1 (\epsilon_1) \theta_1 (\epsilon_2) \theta_1 (2\epsilon_1+ \epsilon_2) \theta_1 (\epsilon_1 + 2\epsilon_2) \theta_i (m \pm \epsilon_+)}  \nonumber\\
  & - \sum_{i=1}^{4} \frac{\eta^{-12} \theta_1(\epsilon_1 + \epsilon_2) \theta_i(\epsilon_1 + \frac{1}{2} \epsilon_2 + 2m) \theta_i(\epsilon_1 + \frac{1}{2} \epsilon_2 ) \prod_{l=1}^{8} \theta_i (m_l \pm \frac{1}{2}\epsilon_1)}{4\theta_1 (\epsilon_1)^{2} \theta_1(\epsilon_2) \theta_{1}(2 \epsilon_1 + \epsilon_2) \theta_1(\epsilon_1 - \epsilon_2) \theta_i (\frac{3}{2}\epsilon_1 + \epsilon_2 + m) \theta_i (\frac{1}{2}\epsilon_1 + m)} \!+\! (\epsilon_1 \leftrightarrow \epsilon_2) \nonumber\\
  &+ \sum_{i=1}^{4} \frac{\eta^{-12} \theta_1 (\epsilon_+ \pm m) \theta_i(\frac{1}{2}\epsilon_1-\epsilon_2 -m) \theta_i(\frac{3\epsilon_1}{2} +m) \prod_{l=1}^{8}\theta_i(m_l \pm \frac{1}{2}\epsilon_1)}{4\theta_1(\epsilon_{1,2})^{2} \theta_1(2 \epsilon_1) \theta_1 (\epsilon_1 - \epsilon_2) \theta_i(\frac{3}{2}\epsilon_1 + \epsilon_2 + m)\theta_i(\frac{1}{2}\epsilon_1 - m)} \,+\, (\epsilon_1 \leftrightarrow \epsilon_2),
\end{align}\end{small}
where residues are arranged in the order in which the poles are listed in. 
There are six $O(2)$ also discrete holonomies whose zero-mode integrals give
\begin{align}
  \mathcal{I}_{(2,1)}^{m} &=  \oint \frac{du}{4} \; \frac{\eta^4 \, \theta_1 (a_1 + a_2) \theta_1 (\epsilon_1 + \epsilon_2 + a_1 + a_2)}{\theta_1( \epsilon_{1,2})^2 \theta_1 (\epsilon_{1,2} + a_1 + a_2)} \frac{\eta^3 \theta_1 (\epsilon_1 + \epsilon_2)}{\theta_1( \epsilon_1)\theta_1( \epsilon_2)} \Bigg(\prod_{l=1}^{8}\prod_{j=1}^2\frac{\theta_1(m_l + a_j)}{ \eta}\Bigg) \nonumber \\
  &\hspace{1.35in}\times \frac{\theta_1(\pm m + \epsilon_- \pm a_1 \mp u)}{\theta_1(\pm m - \epsilon_+ \pm a_1 \mp u)} \frac{\theta_1(\pm m + \epsilon_- \pm a_2 \mp u)}{\theta_1(\pm m - \epsilon_+ \pm a_2 \mp u)}  \frac{\theta_1(m \pm u)}{\theta_1(\epsilon_+ \pm u)}\nonumber\\
  &=\, \frac{\eta^{7} \theta_1 (\epsilon_1 + \epsilon_2) \theta_1 (a_1 + a_2) \theta_1 (\epsilon_1 + \epsilon_2 + a_1 + a_2)}{4\theta_1(\epsilon_{1,2})^3 \theta_1 (\epsilon_{1,2} + a_1 + a_2)}   \Bigg(\prod_{l=1}^{8}\frac{\theta_1 (m_l + a_1) \theta_1 (m_l + a_2)}{\eta^2}\Bigg) \nonumber\\
  \times& \Bigg[\frac{\theta_1(m + \epsilon_+)\theta_1(m - \epsilon_+)}{\eta^3\theta_1(\epsilon_1 + \epsilon_2) } \frac{\theta_1(m + \epsilon_1 +a_{1,2})\theta_1(m + \epsilon_2 + a_{1,2})}{\theta_1(m+a_{1,2})\theta_1(m + \epsilon_1 + \epsilon_2+a_{1,2})} \nonumber \\
  +& \frac{\theta_1 (\epsilon_1)\theta_1 (\epsilon_2)}{\eta^3 \theta_1 (\epsilon_1 + \epsilon_2)} \frac{\theta_1 (\epsilon_1 + a_1 - a_2)\theta_1 (\epsilon_2 + a_1 - a_2)}{\theta_1(a_1 - a_2) \theta_1 (\epsilon_1 + \epsilon_2 + a_1 - a_2)}\frac{\theta_1 (2m + \epsilon_+ + a_1)\theta_1 (\epsilon_+ + a_1)}{\theta_1 (m + \epsilon_1 + \epsilon_2 + a_1)\theta_1 (m + a_1)}\nonumber\\
  +& \frac{\theta_1 (\epsilon_1)\theta_1 (\epsilon_2)}{\eta^3 \theta_1 (\epsilon_1 + \epsilon_2)} \frac{\theta_1 (\epsilon_1 + a_2 - a_1)\theta_1 (\epsilon_2 + a_2 - a_1)}{\theta_1(a_2 - a_1) \theta_1 (\epsilon_1 + \epsilon_2 + a_2 - a_1)}\frac{\theta_1 (2m + \epsilon_+ + a_2)\theta_1 (\epsilon_+ + a_2)}{\theta_1 (m + \epsilon_1 + \epsilon_2 + a_2)\theta_1 (m + a_2)}\Bigg].
\end{align}
where $(a_1, a_2)=(0, \frac{1}{2}), (\frac{\tau}{2},\frac{1+\tau}{2}), (0,\frac{\tau}{2}),(\frac{1}{2},\frac{1+\tau}{2}),(0,\frac{1+\tau}{2}),(\frac{1}{2},\frac{\tau}{2})$ for $m = 1,2,\cdots,6$. Again, this result is in agreement with $ Sp(2) $ instanton calculus.

\subsection{The $(n_1, n_2) = (3,1)$ sector}
This sector contains four rank one integrals and four rank two integrals. The rank one integrals are given by
\begin{align}
  \mathcal{I}_{(3,1)}^{n'} &= \frac{\eta^3\, \theta_1 (2 \epsilon_+)}{\theta_1(\epsilon_{1,2})}\frac{\eta^6}{\theta_1 (\epsilon_{1,2})^3} \left(\prod_{(i,j)} \frac{\theta_1 (a_i + a_j) \theta_1 (2 \epsilon_+ + a_i + a_j)}{\theta_1 ( \epsilon_{1,2}+ a_i + a_j) }\right) \left(\prod_{l=1}^{8} \prod_{i=1}^{3} \frac{\theta_1 (m_l + a_i)}{\eta}\right) \nonumber \\
  & \times \oint \frac{du}{8} \left(\prod_{i=1}^{3} \frac{\theta_1 (\pm m + \epsilon_- \pm a_i \mp u)}{\theta_1 (\pm m - \epsilon_+ \pm a_i \mp u)}\right) \frac{\theta_1 (m \pm u)}{\theta_1 (\epsilon_+ \pm u)},
\end{align}
where $(i,j) \in \{(1,2), (2,3), (3,1)\}$ and $(a_1, a_2, a_3)$ take the following values:
\[\{(\frac{1}{2}, \frac{1+\tau}{2}, \frac{\tau}{2}), (\frac{1}{2}, \frac{1+\tau}{2}, 0), (\frac{\tau}{2}, \frac{1+\tau}{2}, 0), (0, \frac{\tau}{2}, \frac{1}{2})\}.\] Picking up poles at $u = -\epsilon_+, u=m + \epsilon_+ + a_1 , u=m + \epsilon_+ + a_2 , u=m + \epsilon_+ + a_3$, we get
\begin{align}
  \mathcal{I}_{(3,1)}^{n'} &= \frac{\eta^9 \, \theta_1 (2 \epsilon_+)}{8\theta_1 (\epsilon_{1,2})^4} \left(\prod_{(s,t)} \frac{\theta_1 (a_s + a_t) \theta_1 (2 \epsilon_+ + a_s + a_t)}{\theta_1 ( \epsilon_{1,2}+ a_s + a_t) }\right) \left(\prod_{l=1}^{8} \prod_{i=1}^{3} \frac{\theta_1 (m_l + a_i)}{\eta}\right) \nonumber\\
  &\times \Bigg[\frac{\theta_1(m \pm \epsilon_+)}{\eta^3 \, \theta_1 (2 \epsilon_+)} \prod_{i=1}^{3}\frac{\theta_1 (\epsilon_{1,2} + m + a_i)}{\theta_1 (m + a_i) \theta_1 (2 \epsilon_+ +m + a_i )}\,+ \nonumber \\
  & \sum_{(i,j,k)}\frac{\theta_1 (\epsilon_{1,2})}{\eta^3 \, \theta_1 (2 \epsilon_+)} \frac{\theta_1 (\epsilon_+ + a_i) \theta_1 (\epsilon_+ +2m + a_i)}{\theta_1 (m + a_i) \theta_1 (2 \epsilon_+ + m + a_i)} \frac{\theta_1 (\epsilon_{1} + a_i - a_{j,k}) \theta_1 (\epsilon_{2} + a_i - a_{j,k})}{\theta_1 (a_i - a_{j,k})  \theta_1 (2\epsilon_+ + a_i - a_{j,k})}\Bigg]
\end{align}
for $(s,t) \in \{(1,2), (2,3), (3,1)\}$ and $(i,j,k) \in \{(1,2,3), (2,3,1), (3,1,2)\}$. The rank two integrals are:
\begin{align}
  \mathcal{I}_{(3,1)}^{n} &= \oint \frac{du_1 du_2}{4} \frac{\theta_1 (2 \epsilon_+)^2 \theta_1 (2 \epsilon_+ \pm u_1 + a_0) \theta_1 (\pm u_1 + a_0)}{\theta_1 (\epsilon_{1,2})} \prod_{l=1}^{8}\frac{\theta_1 (m_l \pm u_1) \theta_1 (m_l + a_0)}{\eta^3}\nonumber \\
  &\hspace{.15in} \times \frac{\eta^6} {\theta_1(\epsilon_{1,2})^2 \theta_1(\epsilon_{1,2} \pm u_1 + a_0) \theta_1 (\epsilon_{1,2} \pm 2 u_1)} \frac{\theta_1 (\pm m + \epsilon_- \pm a_0 \mp u_2)}{\theta_1 (\pm m - \epsilon_+ \pm a_0 \mp u_2)} \frac{\theta_1 (m \pm u_2)}{\theta_1 (\epsilon_+ \pm u_2)}\nonumber\\
  &\hspace{.15in}\times  \frac{\theta_1 (\pm m + \epsilon_- +u_1 \mp u_2)}{\theta_1 (\pm m - \epsilon_+ +u_1  \mp u_2)}  \frac{\theta_1 (\pm m + \epsilon_- - u_1 \mp u_2)}{\theta_1 (\pm m - \epsilon_+ -u_1 \mp u_2)}, \nonumber
\end{align}
where $a_0$ takes the value $(0, \frac{1}{2}, \frac{1+\tau}{2}, \frac{\tau}{2})$ for $n = 1,\dots,4$. If we choose $\eta = e_1 + \epsilon \, e_2$ with $\epsilon \ll 1$, the following poles are picked by the Jeffrey-Kirwan prescription:
\begin{itemize}
  \item $u_1 + u_2 - \epsilon_+ - m =0$, $\epsilon_+ - u_2 = 0$: $(u_1, u_2) = (m , \epsilon_+)$.
  \item $u_1 + u_2 - \epsilon_+ - m =0$, $m - \epsilon_+ + a_0 - u_2= 0$: $(u_1, u_2) = (2\epsilon_+ - a_0, m - \epsilon_+ + a_0)$.\\
  The corresponding residue does vanish because of $\theta_1 (2 \epsilon_+ \pm u_1 +a_0)$ in the numerator.
  \item $u_1 - u_2 - \epsilon_+ + m = 0$, $\epsilon_+  + u_2 = 0$: $(u_1 , u_2) = (-m, -\epsilon_+)$.
  \item $u_1 - u_2 - \epsilon_+ + m =0$, $- m - \epsilon_+ - a_0  +u_2 = 0$: $(u_1, u_2) = (2\epsilon_+ + a_0, m + \epsilon_+ + a_0)$.\\
  The corresponding residue does vanish because of $\theta_1 (2 \epsilon_+ \pm u_1 +a_0)$ in the numerator.
  \item $u_1 + u_2 - \epsilon_+ - m =0$, $u_1 - u_2 - \epsilon_+ + m = 0$: $(u_1, u_2) = (\epsilon_+ + a_i, m + a_i)$.
  \item $u_1 + u_2 - \epsilon_+ - m = 0$, $2 u_1 + \epsilon_{1,2} = 0$: $(u_1, u_2) = (-\tfrac{\epsilon_{1,2}}{2}+a_i, \epsilon_{1,2}+\tfrac{\epsilon_{2,1}}{2}+m-a_i)$.
  \item $u_1 + u_2 - \epsilon_+ - m = 0$, $u_1 + \epsilon_{1,2} + a_0 = 0$: $(u_1, u_2) = (-\epsilon_{1,2}-a_0, \frac{3\epsilon_{1,2}}{2} + \frac{\epsilon_{2,1}}{2} + m + a_0)$.
  \item $-u_1 + u_2 - \epsilon_+ - m = 0$, $2 u_1 + \epsilon_{1,2} = 0$: $(u_1, u_2) = (-\tfrac{\epsilon_{1,2}}{2}+a_i, \frac{\epsilon_{2,1}}{2}+m+a_i)$.\\
  The corresponding residue does vanish because of $\theta_1 ( \mp m + \epsilon_- \pm u_1 \pm  u_2)$ in the numerator.
  \item $-u_1 + u_2 - \epsilon_+ - m = 0$, $u_1 + \epsilon_{1,2} + a_0 = 0$: $(u_1, u_2) = (-\epsilon_{1,2}-a_0, -\frac{\epsilon_{1,2}}{2} + \frac{\epsilon_{2,1}}{2} - a_0 + m)$.\\
  The corresponding residue does vanish because of $\theta_1 ( \pm m + \epsilon_- \pm a_0 \mp  u_2)$ in the numerator.
  \item $2 u_1 + \epsilon_{1,2} = 0$, $u_2 + \epsilon_+ = 0$: $(u_1, u_2) = (-\frac{\epsilon_{1,2}}{2} + a_i, -\epsilon_+)$.
  \item $2 u_1 + \epsilon_{1,2} = 0$, $u_2 -m - \epsilon_+ - a_0 = 0$: $(u_1, u_2) = (-\frac{\epsilon_{1,2}}{2} + a_i, m + \epsilon_+ + a_0)$.
  \item $u_1 + \epsilon_{1,2} + a_0 = 0$, $u_2 + \epsilon_+ = 0$: $(u_1, u_2) = (- \epsilon_{1,2} -a_0, -\epsilon_+)$.
  \item $u_1 + \epsilon_{1,2} + a_0 = 0$, $u_2 -m - \epsilon_+ - a_0 = 0$: $(u_1, u_2) = (-\epsilon_{1,2} - a_0, m + \epsilon_+ + a_0)$.\\
  The corresponding residue does vanish because of $\theta_1 ( \mp m + \epsilon_- \pm u_1 \pm  u_2)$ in the numerator.
\end{itemize}
In the above, $a_i$ runs over $\{0, \frac{1}{2}, \frac{1+\tau}{2}, \frac{\tau}{2}\}$ while $a_0$ is fixed (dependent on $n$). Collecting the contributions from all residues, one gets:

\begin{align*}
  \mathcal{I}_{(3,1)} &=-\frac{\eta^{-18} \, \theta_1 (\epsilon_+ \pm m)  \theta_n (2 \epsilon_+ + m) \theta_n (m) \prod_{l=1}^{8}\theta_1 (m_l \pm m) \theta_n (m_l)}{4\theta_1(\epsilon_{1,2})^2 \theta_1 (2 \epsilon_+ - 2m) \theta_1 (2m) \theta_1 (\epsilon_{1,2} + 2m) \theta_n (\epsilon_{1,2}+m)}\\
  &+\frac{ \eta^{-18}\, \theta_1 (\epsilon_+ \pm m)  \theta_n (2 \epsilon_+ - m) \theta_n (m) \prod_{l=1}^{8}\theta_1 (m_l \pm m) \theta_n (m_l)}{4\theta_1(\epsilon_{1,2})^2 \theta_1 (2 \epsilon_+ + 2m) \theta_1 (2m) \theta_1 (\epsilon_{1,2} - 2m) \theta_n (\epsilon_{1,2}-m)}\\    
  & - \sum_{(i,j) \in S_n} \frac{\eta^{-18}\, \theta_i (0) \theta_i (2m) \theta_j (\epsilon_+) \theta_j (3 \epsilon_+) \prod_{l=1}^{8} \theta_i (m_l \pm \epsilon_+ ) \theta_n (m_l)}{8\theta_1(\epsilon_{1,2})^2  \theta_i (m \pm \epsilon_+) \theta_j (\frac{3 \epsilon_{1,2}}{2} + \frac{\epsilon_{2,1}}{2})  \theta_1 (2 \epsilon_{1,2} + \epsilon_{2,1})} \\
  & + \Bigg[ \sum_{(i,j) \in S_n} \frac{\eta^{-18} \theta_1 (2 \epsilon_+) \theta_i (\epsilon_1 \!+\! \frac{\epsilon_2}{2} ) \theta_i (\epsilon_1 \!+\! \frac{\epsilon_2}{2}\!+\!2m) \theta_j (\frac{\epsilon_1}{2}\!+\!\epsilon_2)  \prod_{l=1}^{8} \theta_i (m_l \!\pm\! \frac{\epsilon_1}{2}) \theta_n (m_l)}{8\theta_1 (\epsilon_1)^3 \theta_1 (\epsilon_2)^2 \theta_1 (\epsilon_1 \!-\! \epsilon_2) \theta_1 (2 \epsilon_1 \!+\! \epsilon_2) \theta_i (\frac{\epsilon_1}{2} \!+\! m) \theta_i (\frac{3 \epsilon_1}{2} \!+\! \epsilon_2 + m) \theta_j (\frac{\epsilon_1}{2} \!-\! \epsilon_2)}\\
  & - \frac{\eta^{-18} \theta_1 (2 \epsilon_+) \theta_n (\frac{3 \epsilon_1}{2}+ \frac{\epsilon_2}{2}+2m ) \theta_n (\frac{3 \epsilon_1}{2}+ \frac{\epsilon_2}{2} ) \prod_{l=1}^{8} \theta_n ( m_l \pm \epsilon_1 ) \theta_n (m_l )}{4\theta_1 (\epsilon_1)^2 \theta_1 (2 \epsilon_1) \theta_1 ( \epsilon_1 \!-\! \epsilon_2) \theta_1 (2 \epsilon_1 \!-\! \epsilon_2) \theta_1 (\epsilon_2) \theta_1 (3 \epsilon_1 \!+\! \epsilon_2) \theta_n (\epsilon_1 \!+\! m ) \theta_n (2 \epsilon_1 \!+\! \epsilon_2 + m )} \\
  & - \sum_{(i,j)\in S_n}  \Bigg(\frac{\eta^{-18} \theta_1 (\epsilon_+ \pm m) \theta_n (\epsilon_{1,2} + m ) \theta_i (\frac{\epsilon_1}{2}- \epsilon_2-m ) \theta_i (\frac{3 \epsilon_1}{2} + m ) }{8\theta_1 (\epsilon_{1,2})^3  \theta_1 (2 \epsilon_1) \theta_1 (\epsilon_1 - \epsilon_2) \theta_n (m) \theta_n ( \epsilon_1 + \epsilon_2 + m) }  \\
  &\hspace{2.1in}\times \frac{\theta_j (\frac{\epsilon_1}{2}) \theta_j (\frac{3 \epsilon_1}{2}+ \epsilon_2) \prod_{l=1}^{8} \theta_i (m_l \pm \frac{\epsilon_1}{2}) \theta_n (m_l) }{\theta_i (\frac{\epsilon_1}{2}-m) \theta_i (\frac{3 \epsilon_1}{2} + \epsilon_2 + m) \theta_j(\frac{\epsilon_1}{2} - \epsilon_2) \theta_j (\frac{3 \epsilon_1}{2})}\Bigg)\nonumber\\
  & - \sum_{(i,j) \in S_n} \frac{\eta^{-18}\theta_n (\epsilon_+) \theta_n (\epsilon_+ + 2m) \prod_{l=1} \theta_i (m_l \pm \frac{\epsilon_1}{2}) \theta_n(m_l)}{8\theta_1 (\epsilon_{1,2})^2 \theta_1(2 \epsilon_1) \theta_1 (\epsilon_1 - \epsilon_2) \theta_n (m) \theta_n (\epsilon_1 + \epsilon_2 + m)} \\
  & - \!\frac{\eta^{-18} \theta_1 (m \!\pm\! \epsilon_+) \theta_n (\epsilon_1\! -\! \epsilon_2 \!-\! m ) \theta_n (2 \epsilon_1 \!+\! m) \prod_{l=1}^{8} \theta_n (m_l \pm \epsilon_1) \theta_n(m_l) }{4\theta_1\!(\epsilon_{1,2})^2 \theta_1\! (2 \epsilon_1) \theta_1\! (3 \epsilon_1) \theta_1\! (\epsilon_1 \!-\! \epsilon_2) \theta_1\! (2 \epsilon_1 \!-\! \epsilon_2) \theta_n\! (\epsilon_1\!-\! m ) \theta_n (2 \epsilon_1 \!+\! \epsilon_2 \!+\! m )}
\!+\! (\epsilon_1\! \leftrightarrow\! \epsilon_2) \Bigg],
\end{align*}
where terms are arranged in order which poles are listed in. In the above,
 \begin{align*} S_1 &= \{(1,1),(2,2),(3,3),(4,4)\},\\
 S_2 &= \{ (1,2), (2,1), (3,4), (4,3)\},\\
 S_3 &= \{(1,3),(2,4),(3,1),(4,2)\},\\
 S_4 &= \{(1,4),(2,3),(3,2),(4,1)\}.
 \end{align*}
  We have checked that this computation also agrees with $ Sp(2) $ instanton calculus.

\bibliography{references}

\end{document}